\documentclass[12pt]{article}

\usepackage{authblk}

\usepackage{amsmath}
\usepackage{graphicx}
\usepackage{xcolor} 
\usepackage{caption}
\usepackage{amssymb} 
\usepackage{comment}
\usepackage[backend=biber, style=ieee, citestyle=authoryear]{biblatex}
\let\cite\parencite
\usepackage{hyperref} 
\hypersetup{
    colorlinks=true,
    linkcolor=blue,       
    citecolor=blue,          
    urlcolor=blue,    
}
\title{Effects of Periodic Migration on Selection in Subdivided Populations}

\author[1]{Mohammad Amin Masoumi\thanks{Email: \href{mailto:massomiamin@gmail.com}{massomiamin@gmail.com}}}
\author[2]{Kamran Kaveh\thanks{Email: \href{mailto:kkavehma@gmail.com}{kkavehma@gmail.com}}}
\author[1,3]{Mohammad Reza Ejtehadi\thanks{Email: \href{mailto:ejtehadi@sharif.edu}{ejtehadi@sharif.edu}}}

\affil[1]{Department of Physics, Sharif University of Technology, Tehran, Iran}
\affil[2]{University of Minnesota, Minneapolis, USA}
\affil[3]{ICTP – The Abdus Salam International Centre for Theoretical Physics, 34151 Trieste, Italy}

\date{\today}

\begin{document}

\maketitle
\begin{abstract}
Migration and selection jointly shape evolutionary outcomes in spatially structured populations. Migration can vary over time as ecological shifts, such as changes in sea level, alter habitat connectivity. Here, we study the impact of periodic migration on selection within a geographically subdivided population. Using a two-deme Moran model with spatially heterogeneous selection, we analyze a scenario in which a mutant genotype is advantageous in one habitat and disadvantageous in the other. Keeping the time-averaged migration rate fixed, we vary the period, relative phase, and directional asymmetry of migration and examine the resulting steady-state mutant frequency. We find that migration periodicity generally drives the population toward neutrality: relative to constant migration, it decreases the mutant frequency when mutants are advantageous on average and increases it when they are deleterious, without reversing the direction of selection in the parameter regimes examined. This effect is strongest for out-of-phase migration, which models seasonal migration, and is further enhanced by directional asymmetry. The response to the migration period is generally non-monotonic, with an intermediate characteristic period producing the largest deviation from constant-migration behavior under in-phase migration. These results arise from the coupling between migration and local selection in Moran process: temporal modulation of migration changes not only the exchange of individuals between habitats but also the local population composition on which selection acts. Our results show that populations with identical mean migration rates can show different evolutionary outcomes depending on the temporal organization of gene flow.
 \end{abstract}
\section{Introduction}

The fate of a new mutant strategy is shaped not only by its intrinsic fitness advantage but also by the demographic and spatial structure in
which competition takes place \cite{nowak2006evolutionary, broom2014game, nagylaki1992introduction,durrett2008probability}.
Population structure, spatial heterogeneity in the selective environment, and temporal variation in ecological conditions can each
promote, suppress, or reverse the success of a type that would be unambiguously favored or disfavored in a well-mixed population 
\cite{maruyama1970fixation, nagylaki1980strong,maruyama1, levins1966maintenance, levene1953genetic, gavrilets2002fixation, blanquart2012effects, kaveh2019environmental, svoboda2025gradient, campos2008environmental, farhang2017effect,gorgi2025spatio}.

Beyond inherent fitness, one trait that plays a particularly consequential role in structured populations is motility: the capacity of individuals or their offspring
to move between distinct habitats or demes. Motility governs the pattern of gene flow between subpopulations, and its rate,
directionality, and temporal structure act as a filter through which local selective differences are expressed at the population level,
whether or not motility itself is type-dependent \cite{levene1953genetic, nagylaki1980strong, lenormand2002gene,
krieger2017effects}. In heterogeneous environments, where the same genotype may be favored in one habitat and disfavored in another, the
interplay between migration and local selection is often decisive: gene flow can rescue populations from local maladaptation, or it can
overwhelm local selection and prevent the maintenance of polymorphism \cite{levene1953genetic,levins1966maintenance,tachida1991fixation,gavrilets2002fixation}.

The interplay between migration and local selection has been a central problem in evolutionary theory since its foundations.
Haldane~\cite{haldane1930mathematical} showed that a locally favored allele is lost from an island population once the immigration rate exceeds the local selection coefficient, establishing a sharp threshold above which gene flow overwhelms local adaptation.
Levene~\cite{levene1953genetic} demonstrated the complementary result: spatially heterogeneous selection across multiple niches can maintain stable polymorphism even under complete dispersal, provided the selective differences between habitats are large enough. A unifying conclusion from this body of work \cite{wright1931evolution, karlin1980selection, bulmer1972multiple, lenormand2002gene} is that evolutionary outcomes in structured populations depend sensitively on the \emph{ratio} of migration rate to selection coefficient: below a critical threshold, locally adapted alleles are protected; above it, gene flow homogenizes the population and polymorphism collapses.

Natural populations rarely experience constant selective environments. Temporal fluctuations in resource availability, temperature, predation pressure, and other ecological variables cause fitness to vary over time, sometimes reversing in sign across seasons or generations.
Dempster~\cite{dempster1955maintenance} and Haldane and Jayakar~\cite{haldane1963polymorphism_varying} showed that in a
panmictic population under temporally varying selection, the fate of a polymorphism is governed by the \emph{geometric} mean of fitness across environments, not its arithmetic mean, so that temporal variance in fitness, and not merely its average, shapes evolutionary outcomes. Subsequent work extended this insight to structured populations, showing that fluctuating selection interacts with migration and deme
size in complex, often non-intuitive ways \cite{gillespie1973polymorphism, gillespie1976role, felsenstein1976theoretical, hedrick1976genetic, hedrick2006genetic,
farhang2019environmental, ashcroft2014fixation,kaveh2020moran}. Most notably, Cvijovi\'c et al.~\cite{cvijovic2015fate} showed that temporal
fluctuations in selection can substantially reduce the efficiency of selection and, under sufficient variability, even reverse the fate of
a mutation relative to its time-averaged fitness. A recurring theme across this literature is that temporal variation in selective
pressures can either promote or erode polymorphism depending on its interaction with spatial structure and gene flow. In nearly all of
these models, however, the migration rate itself is held constant.

Beyond fitness, migration itself is also subject to temporal variation. Seasonal cycles, fluctuating geographic barriers, and
episodic dispersal events all impose temporal structure on the rate and directionality of gene flow between populations. Migratory
species move between breeding and non-breeding habitats on annual schedule~\cite{peniston2019pulsed}; land bridges
form and submerge with sea-level oscillations~\cite{princepe2024intermittent}; and pollinators, insect
vectors, and pathogens disperse between habitats on seasonal timescales. Despite the ubiquity of these phenomena, the theoretical
consequences of temporally structured migration have received far less attention than the analogous case of temporally varying selection \cite{krieger2017effects,rice2009evolution,aubree2023migration}.
Existing work has largely focused on intermittent or pulsed migration in mainland--island geometries, where episodic connectivity between
otherwise isolated populations generates bursts of gene flow, and has shown that even in the absence of selection such pulses can produce
non-trivial evolutionary signatures~\cite{princepe2024intermittent, peniston2019pulsed,aubree2023migration}. Comparatively little is known, however, about how smooth, periodic
migration, as would arise most naturally from seasonal environmental cycles, interacts with spatially heterogeneous selection to shape
the steady-state balance between local adaptation and gene flow.

Figure~\ref{fig:Moran1} illustrates two habitats separated by a seasonally varying barrier (panel a), across which migration occurs at time-dependent rates \(\mu_{12}(t)\) and \(\mu_{21}(t)\) (panel b). In habitat 1, type A is locally favored (\(r_{A,1}=r_1>r_B=1\)), while in habitat 2 type B is favored (\(r_{A,2}=r_2<r_B=1\)).  

This situation is realized in coastal and island systems, where tides and seasonal currents generate intermittent connectivity over short to long timescales. The same structure appears in agricultural and epidemiological settings, in which vectors or pathogens move seasonally between natural refuges and cultivated habitats that impose strong selection on resistance alleles. In both classes of systems, the temporal schedule of migration itself, not merely the total amount of gene flow, determines whether locally favored traits are protected or eroded.

% ---------- Paragraph 6: model ----------

Here we address this gap using a two-deme Moran death--birth process in
which each deme has a fixed but distinct selective environment: a
mutant type is favored in one habitat and disfavored in the other.
Migration between demes is bidirectional and periodic, with
a choice of in-phase or out-of-phase migration , and directional asymmetry. Deme sizes are taken large
enough that the dynamics are well approximated by deterministic
mean-field equations, isolating the effect of temporal migration
structure from stochastic fluctuations. Throughout, we hold the
time-averaged migration rate fixed and ask how the temporal pattern
of gene flow, its period, phase, and symmetry, reshapes the
steady-state balance between local selection and migration.

% ---------- Paragraph 7: findings ----------

Our central finding is that periodic migration modulates the
steady-state mutant frequency $\bar\phi$ in a \emph{sign-preserving}
manner. When the mutant is advantageous on average
($\bar\phi > 0.5$), periodic migration reduces $\bar\phi$ relative to
constant migration; when the mutant is deleterious on average
($\bar\phi < 0.5$), periodic migration increases $\bar\phi$. In neither
case does the modulation cross the $\bar\phi = 0.5$ neutrality line.
Temporal migration structure thus acts as a \emph{contractor toward
neutrality}: it attenuates the effective strength of selection without
inverting its direction, in sharp contrast to the reversal effects
that fluctuating selection can produce~\cite{cvijovic2015fate}. The
magnitude of the contraction is strongest under out-of-phase migration
and is markedly amplified when migration is asymmetric between demes.
The dependence on the migration period $T$ is generically
non-monotonic, with an intermediate $T_{\text{opt}}$ at which the
effect is extremal, suggesting a resonance between the migration
timescale and the intrinsic relaxation timescale of local selection.

% ---------- Paragraph 8: roadmap ----------

The remainder of the paper is organized as follows.
Section~\ref{sec:model} defines the model and the migration scenarios
we consider. Section~\ref{sec:results} presents the results for
symmetric and asymmetric periodic migration under the three selective
regimes. Section~\ref{sec:discussion} discusses the implications and
limitations of our findings.

\section{Model}
\label{sec:model}

We employ the Moran process~\cite{moran} to model evolutionary dynamics in a finite structured population. The population consists
of two demes (``islands'') of equal, constant size $N$, occupied by two types of individuals: mutants (type $A$) and wild-type residents
(type $B$). The two demes differ in resource availability, such that mutants have island-specific birth rates $r_{A1}$ on island~1 and
$r_{A2}$ on island~2, while residents have a birth rate $r_B$ on both islands.

Migration between demes is encoded in a time-dependent migration matrix $\boldsymbol{\mu}(t) = \bigl(\mu_{ij}(t)\bigr)$, where
$\mu_{ij}(t)$ is the probability that an offspring produced on island~$j$ migrates to island~$i$ at time~$t$. In the two-deme case
the free parameters are the off-diagonal probabilities $\mu_{12}(t)$ and $\mu_{21}(t)$, with retention probabilities fixed
by the normalization $\mu_{i1}(t) + \mu_{i2}(t) = 1$. Sections~\ref{sec:symmetric} and~\ref{sec:asymmetric} introduce the particular periodic choices of $\mu_{12}(t)$ and $\mu_{21}(t)$ that characterize the symmetric, asymmetric, in-phase, and out-of-phase cases examined in this work.

\begin{figure}[h]
\centering
\includegraphics[width=0.9\textwidth]{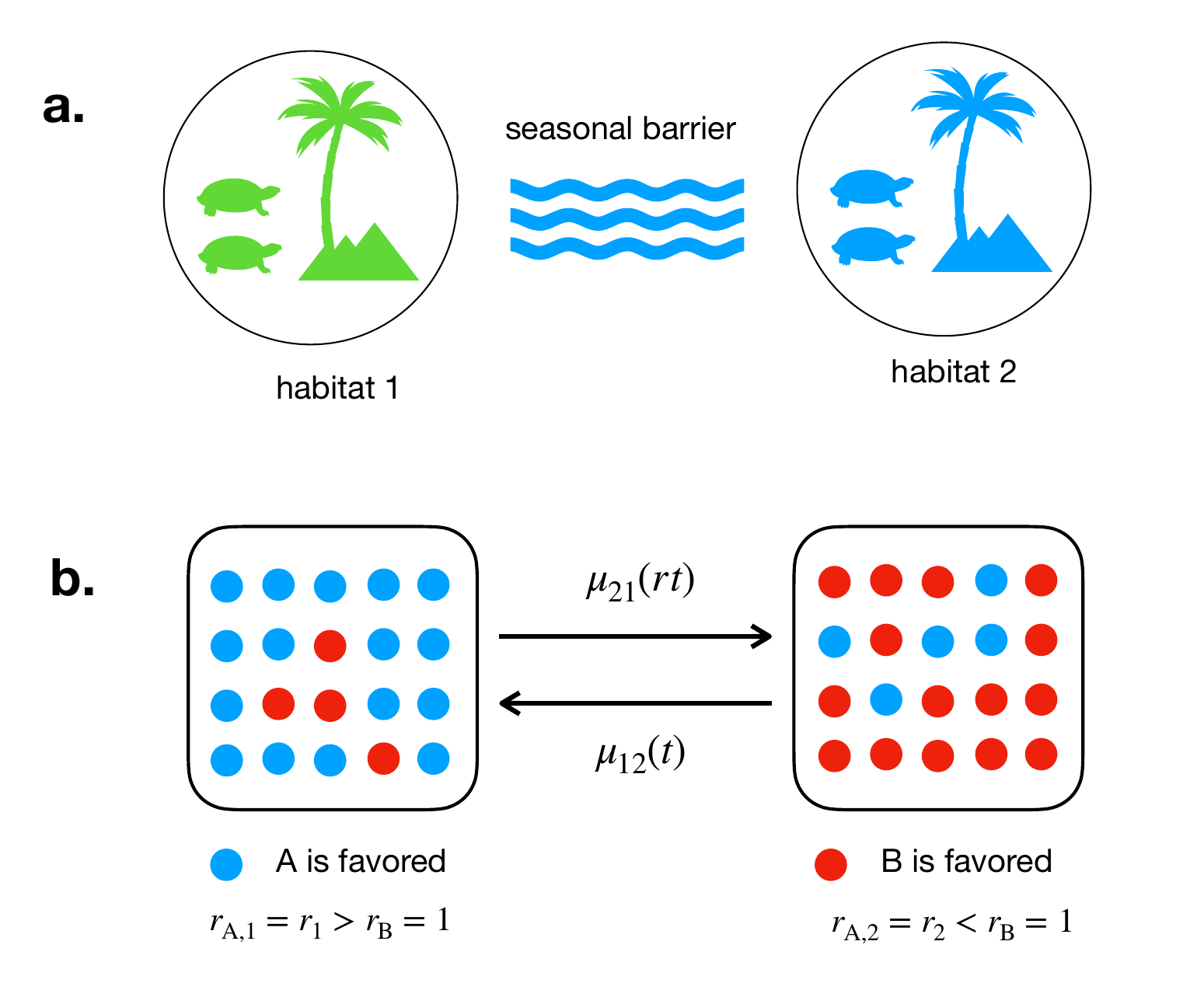}
\caption{}
% \caption{\textbf{Schematic representation of the model.}
% \textbf{(A)} Illustration/example of two geographically separated habitats, by water or sea, where
% temporal environmental variation, such as seasonal fluctuations in sea level
% can produce time-dependent migration between the populations.
% \textbf{(B)} Two-deme population model used in this study. Each habitat
% contains $N$ individuals, mix of mutant type $A$ (blue) and resident type $B$ (red). 
% The mutant has habitat-dependent birth rates $r_{A,1}$ and $r_{A,2}$, whereas the resident birth rate is $r_B=1$ in
% both habitats. Individuals migrate between habitats with time-dependent
% rates $\mu_{12}(t)$ and $\mu_{21}(t)$.}
% % \captionsetup[figure]{labelsep=none}
% % \caption{}
\label{fig:Moran1}
\end{figure}

The dynamics follow a death-birth update rule. At each step, an individual on a focal island, $i$, is selected uniformly at random and
removed. A parent is then chosen with a probability proportional to its fitness from either the focal island itself (with probability
$\mu_{ii}$) or the other island (with probability $\mu_{ij}$), and its offspring occupies the vacated site. The corresponding
transition probabilities for the number of mutants $n$ on the focal island, $W^+(n) = W(n \to n+1)$ and $W^-(n) = W(n \to n-1)$, are
derived in Appendix~\ref{app:transitions}. For deme sizes large enough that stochastic fluctuations are negligible, the dynamics
reduce to a pair of coupled deterministic ordinary differential equations for the mutant frequencies $\phi_i(t) = n_i(t)/N$; the
derivation is given in Appendix~\ref{app:meanfield}, and this mean-field description is used throughout the main text. A schematic
of the model is shown in Figure~\ref{fig:Moran1}.
   
% ---------- Immediately after the schematic figure ----------

\subsection{Periodic migration scenarios}
\label{sec:scenarios}

Seasonal migration in natural populations is often approximately
rhythmic, driven by cyclic environmental cues such as photoperiod,
temperature, and resource availability. To capture this
periodicity in its simplest form, we take the migration rates to be
smooth periodic functions of time with a common period~$T$, and
consider four scenarios that separately vary the \emph{symmetry}
between directions (equal vs.\ unequal amplitudes) and the
\emph{phase relationship} between the two directional flows (in-phase
vs.\ out-of-phase). The four scenarios are summarized in
Table~\ref{tab:scenarios}.

\begin{table}[h]
\centering
\begin{tabular}{clcc}
\hline
 & Scenario & \( \mu_{21}(t) \) & \( \mu_{12}(t) \) \\
\hline
A & Symmetric in-phase       & \( \alpha_0 \cos^2(\pi t/T) \) & \( \alpha_0 \cos^2(\pi t/T) \) \\
B & Symmetric out-of-phase   & \( \alpha_0 \cos^2(\pi t/T) \) & \( \alpha_0 \sin^2(\pi t/T) \) \\
C & Asymmetric in-phase      & \( \alpha_0 \cos^2(\pi t/T) \) & \( \beta_0  \cos^2(\pi t/T) \) \\
D & Asymmetric out-of-phase  & \( \alpha_0 \cos^2(\pi t/T) \) & \( \beta_0  \sin^2(\pi t/T) \) \\
\hline
\end{tabular}
\caption{The four periodic migration scenarios considered in this
study. Amplitudes \( \alpha_0, \beta_0 \in [0,1] \) and period \( T > 0 \)
are the free parameters. Retention probabilities on each island
follow from normalization: \( \mu_{11}(t) = 1 - \mu_{21}(t) \) and
\( \mu_{22}(t) = 1 - \mu_{12}(t) \).}
\label{tab:scenarios}
\end{table}

\begin{figure}[htbp]
\includegraphics[width=0.9\textwidth]{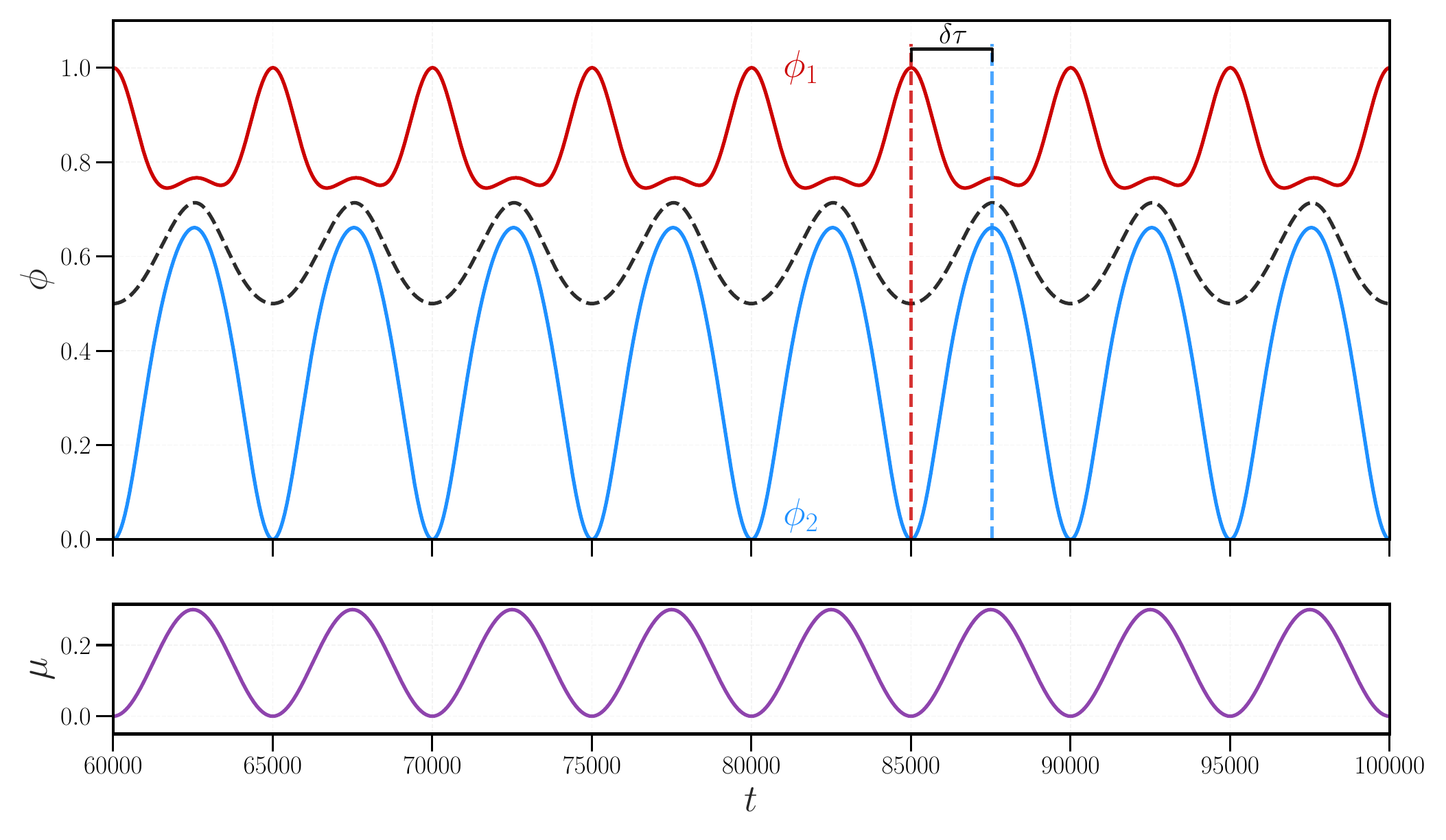}
\caption{Time evolution of the mutant frequencies \(\phi_1\) and \(\phi_2\) (top panel) after an initial transient of 60,000 time units, under periodic migration \(\mu(t)\) (bottom panel). The solid red line denotes \(\phi_1\) (island 1), while the solid blue line represents \(\phi_2\) (island 2). The black dashed line indicates the instantaneous mean mutant frequency.
Vertical red and blue dashed lines indicate times of the selected maxima of the mutant frequencies on the two islands, with the separation between them quantifying the time delay $\delta \tau$.}
\label{fig:Moran2} 
\end{figure}

In the \emph{symmetric} scenarios, migration probabilities in both
directions share a common amplitude~$\alpha_0$; in the
\emph{asymmetric} scenarios, the two directions have independent
amplitudes $\alpha_0$ and $\beta_0$, which allows one to model
prevailing directional flow (e.g., a dominant seasonal current or
prevailing winds). In the \emph{in-phase} scenarios, peak migration in
the two directions coincides; in the \emph{out-of-phase} scenarios,
peak migration in one direction coincides with minimum migration in
the reverse direction, corresponding to a half-period ($T/2$)
temporal offset. This last scenario provides the closest analog to
alternating seasonal dominance, for example, migratory species
that move predominantly from breeding to non-breeding grounds in one
half of the year and reverse in the other. Exchanging $\sin$ and
$\cos$ or swapping island labels leaves the dynamics invariant, so
these four scenarios exhaust the qualitatively distinct cases.

Under all four scenarios, the \emph{time-averaged} migration rate in
each direction is fixed: $\overline{\mu_{12}} = \overline{\mu_{21}} =
\alpha_0 / 2$ in the symmetric case, and $\overline{\mu_{12}} =
\alpha_0/2$, $\overline{\mu_{21}} = \beta_0/2$ in the asymmetric case.
Any effect of periodicity on the steady-state behavior is therefore
attributable to the \emph{temporal structure} of migration, not to a
change in its mean rate.

In addition to migration, the form of selection can strongly influence the dynamics of the populations. The interplay between migration and selection is particularly interesting, as their combination can give rise to a rich variety of outcomes that neither process alone would produce.

To analyze this interplay, we distinguish three regimes according to the mutant birth rates in the two habitats. Recall that \(r_{A1}\) and \(r_{A2}\) denote the intrinsic birth rates of type \(A\) on islands 1 and 2, respectively, and \(r_B\) is the birth rate of type \(B\). We assume type \(A\) is always advantaged on island 1 (\(r_{A1}>r_B\)) and disadvantaged on island 2 (\(r_{A2}<r_B\)). The regimes differ solely in the mean of the two local birth rates of type \(A\) relative to \(r_B\):

\begin{table}[h]
\centering
\begin{tabular}{cl}
\hline
Regime & Condition \\
\hline
I   & \(\displaystyle\frac{r_{A1}+r_{A2}}{2}=r_B\) \\[1.2ex]
II  & \(\displaystyle\frac{r_{A1}+r_{A2}}{2}<r_B\) \\[1.2ex]
III & \(\displaystyle\frac{r_{A1}+r_{A2}}{2}>r_B\) \\
\hline
\end{tabular}
\caption{The three selection regimes in this study. Regime~I corresponds to a balanced (neutral on average) situation, Regime~II to overall selection against type \(A\), and Regime~III to overall selection in favor of type \(A\). In all three regimes type \(A\) is locally advantaged on island 1 (\(r_{A1}>r_B\)) and locally disadvantaged on island 2 (\(r_{A2}<r_B\)).}
\label{tab:regimes}
\end{table}

These three regimes, together with the four migration scenarios introduced in Table~\ref{tab:scenarios}, allow us to systematically examine how the interaction between periodic migration and spatial heterogeneity in selection depends on both the symmetry and timing of migration and the overall strength and direction of selection.

\subsection{Observable}
\label{sec:observable}

For each scenario as shown in Figure \ref{fig:Moran2} we track the mutant frequencies $\phi_1(t)$ and
$\phi_2(t)$ on the two islands, obtained by numerical integration of
the coupled ODEs derived in Appendix~\ref{app:transitions}. After
transient behavior has decayed, the frequencies settle into a
periodic orbit with period~$T$. We characterize this orbit by the
time-averaged total mutant frequency across the two islands,
\begin{equation}
\label{eq:phibar}
    \bar\phi \;=\; \frac{1}{2T} \int_{t_0}^{t_0 +T} \bigl[\phi_1(t) + \phi_2(t)\bigr]\, dt,
\end{equation}
evaluated over a single period in the steady state, where $t_0$ denotes an arbitrary time within the steady-state region.. 

If $\bar\phi > 1/2$ we consider the mutant to be \emph{advantageous} on average across the metapopulation,
and $\bar\phi < 1/2$ correspond to on average \emph{deleterious} mutant ($\bar\phi = 1/2$
is the neutral case). Under this convention, the constant-migration baseline \(\bar{\phi}^{(0)}\) 
serves as the reference against which the effects of periodic migration are measured. This baseline coincides with the low-period limit (\(T\to 0\)) in which the system responds to the time-averaged rates (half the amplitude, \(\alpha_0/2\)). In the opposite limit of very large periods (\(T\to\infty\)), the migration rates remain nearly constant over long intervals, so the dynamics are controlled by the extreme values of the oscillatory functions (essentially the full amplitude, \(\alpha_0\)). Consequently, the two ends of the \(\bar{\phi}\) curves are quantitatively recovered by constant-migration calculations performed at half-amplitude and full-amplitude rates, respectively. In the figures, we display only the larger-period regime because, in the small-\(T\) limit, the oscillatory results coincide with the constant-migration points already shown; extending the plots to still smaller periods would therefore add no new information. 

\section{Results}
\label{sec:results}

In this section, we examine each of the migration scenarios defined above. For each scenario, we characterize the
steady-state mutant frequency~$\bar\phi$ as a function of the migration period~$T$, amplitude, and phase relationship, under the
three selective regimes defined by conditions~\eqref{equation1}--\eqref{equation3}. We begin with the symmetric case and proceed to
asymmetric migration, highlighting throughout how periodicity reshapes the balance between local selection and gene flow. In each
subsection, we present the scenario--regime combinations that most clearly illustrate the key effects; the remaining combinations,
which exhibit qualitatively similar behavior, are reported in the \textcolor{blue}{Appendix C}.

As a baseline, consider the case of complete isolation: no migration occurs between the two islands. When mutants are advantageous on
island~1 but disadvantageous on island~2, they fixate on island~1 while residents fixate on island~2, yielding $\bar\phi = 0.5$. This
isolation baseline serves as the reference against which all migration effects are measured.

\begin{figure}[htbp]
\hfill
\hspace*{-0.15\textwidth}
\includegraphics[width=1.2\textwidth]{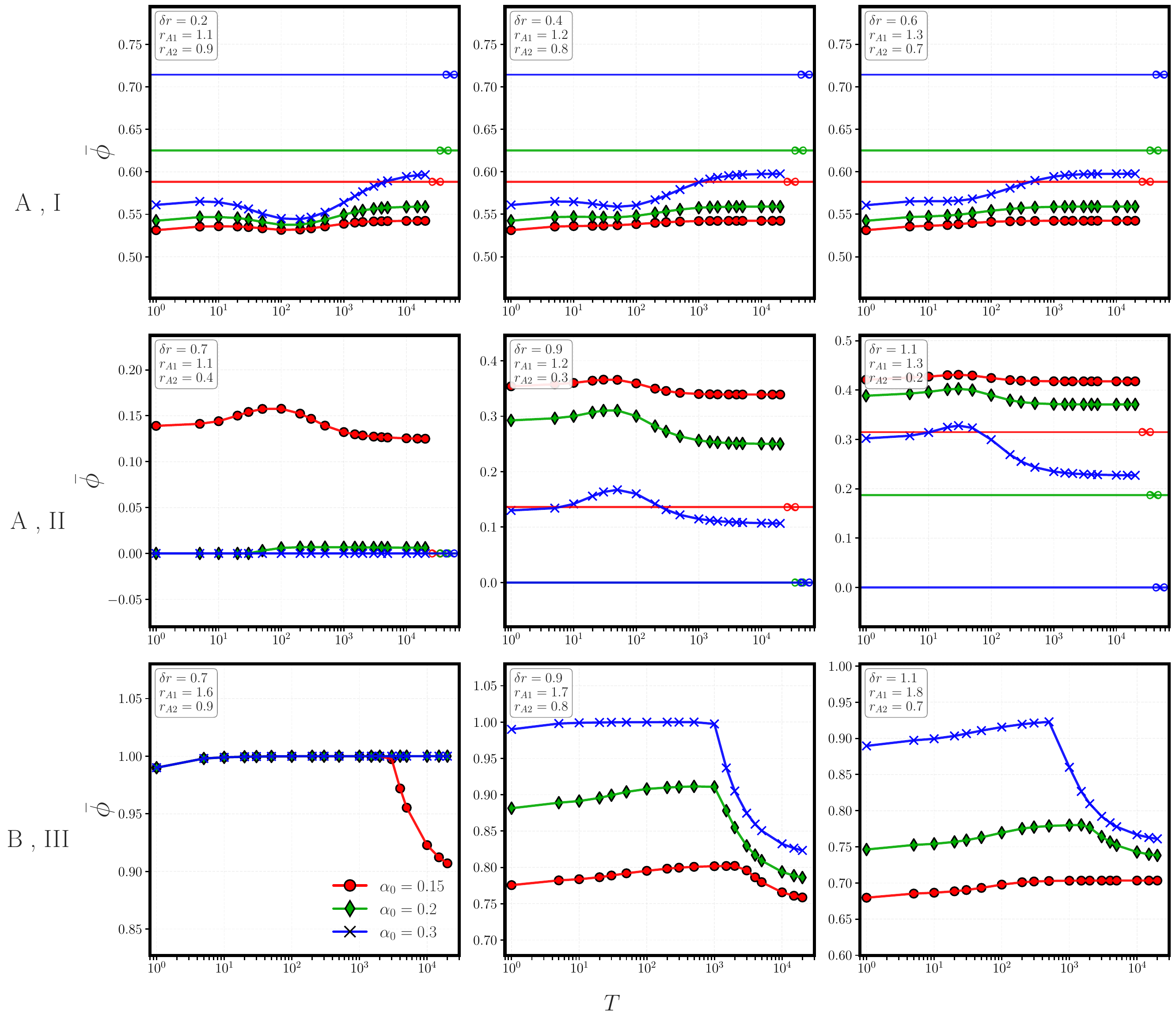}
\caption{Time-averaged mutant frequency~$\bar\phi$ as a function of migration period~$T$ (logarithmic scale) under symmetric periodic
migration. Rows correspond to three representative scenario--regime combinations: (A,\,I)~symmetric in-phase with
condition~\eqref{equation1}, $\delta r = 0.2$, $0.4$, $0.6$; (A,\,II)~symmetric in-phase with condition~\eqref{equation2},
$\delta r = 0.7$, $0.9$, $1.1$; (B,\,III)~symmetric out-of-phase with condition~\eqref{equation3}, $\delta r = 0.7$, $0.9$, $1.1$.
Columns show increasing~$\delta r$ from left to right. In all panels: $\alpha_0 = 0.15$~($\circ$), $\alpha_0 = 0.2$~($\diamond$),
$\alpha_0 = 0.3$~($\times$). Solid lines with open symbols at $T \to \infty$ indicate the large-period limit.} %the dotted horizontal line marks the isolation baseline $\bar\phi = 0.5$.%
\label{fig:Moran3}
\end{figure}

% \begin{figure}[htbp]
% \hfill
% \hspace*{-0.15\textwidth}
% \includegraphics[width=1.2\textwidth]{Fig2.pdf}
% %\includegraphics[width=1.3\textwidth, height=1.5\textheight, keepaspectratio]{Figure7.png}
% \caption{Dynamics of time-averaged mutant frequency ($\bar\phi$) vs period ($T$)(logarithmic scale) in different selection regime and symmetric migration scenarios. 
% (a) Symmetric in-phase with condition \eqref{equation1}: $\delta r = 0.2$ , $0.4$ , $0.6$ . 
% (b) Symmetric in-phase with condition \eqref{equation2}: $\delta r = 0.7$, $0.9$ , $1.1$ . 
% (c) Symmetric out-of-phase with condition \eqref{equation3}: $\delta r = 0.7$ , $0.9$ , $1.1$ . 
% In all panels: $\alpha_0 = 0.15$ ($\circ$), $\alpha_0 = 0.2$ ($\diamond$), $\alpha_0 = 0.3$ ($\times$).}
% \label{fig:Moran3}
% \end{figure}

\subsection{Symmetric Periodic Migration}
\label{sec:symmetric}

We first consider symmetric periodic migration, in which the two
directions share a common amplitude~$\alpha_0$. The simplest
selective regime is neutral selection on average (Table \ref{tab:regimes}):
\begin{equation}\label{equation1}
\tag{I}
\frac{r_{A1} + r_{A2}}{2} = r_B , \quad r_{A1} > r_B , \quad
r_{A2} < r_B,
\end{equation}
so that the mutant is favored on island 1 and disfavored on island 2 by equal and opposite amounts, by this we define fitness difference ~$\delta r = r_{A1} - r_{A2}$ .  
Figure~\ref{fig:Moran3}(A,\,I) shows the steady-state mutant
frequency~$\bar\phi$ as a function of the migration period~$T$ under
in-phase symmetric migration with condition~\eqref{equation1}. The
straight solid lines at $T \to \infty$ indicate the large-period
limit. Solid lines raises~$\bar\phi$ above the isolation
baseline ($\bar\phi = 0.5$), because continuous gene flow allows
advantageous mutants to spread from island~1 into island~2.
Introducing periodicity slightly reduces~$\bar\phi$ relative to this
large-period limit: periodic interruptions temporarily cut off
the flow of migrants, weakening the net effect of gene flow. As~$T$
increases, $\bar\phi$ rises modestly, approaching the
constant-migration value, consistent with the expectation that
longer periods more closely approximate continuous migration. Under
out-of-phase migration (\textcolor{blue}{Appendix C \ref{sec:S1.1} }), the reduction
in~$\bar\phi$ relative to the constant-migration limit is
substantially stronger, and this suppressive effect intensifies with
increasing seasonal period.

\begin{figure}[htbp]
\hfill
\includegraphics[width=1.1\textwidth]{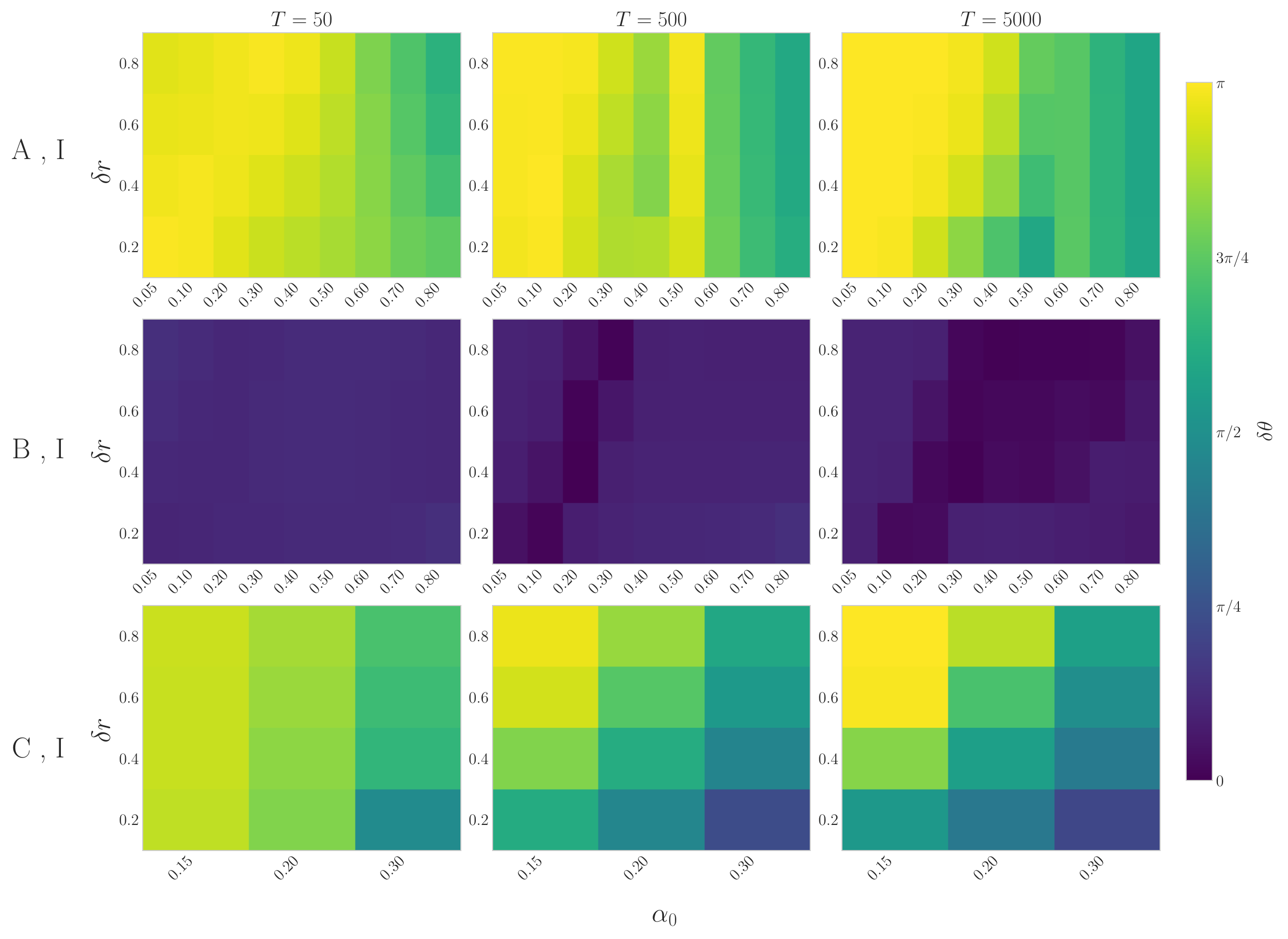}
\caption{Phase difference~$\delta_\theta$ between
mutant-frequency oscillations on the two islands as a function of
migration amplitude~$\alpha_0$ (horizontal axis) and fitness
difference~$\delta r = r_{A1} - r_{A2}$ (vertical axis), shown
under selection regime~\eqref{equation1}. Columns correspond to
three seasonal periods ($T = 50$, $500$, and $5000$); rows
correspond to three migration scenarios (A:~symmetric in-phase,
B:~symmetric out-of-phase, C:~asymmetric in-phase with $\beta_0$ fixed). Color
indicates the magnitude of~$\delta_\theta$ in radians, from
$\delta_\theta \approx 0$ (dark purple) to
$\delta_\theta \approx \pi$ (bright yellow).}
\label{fig:Moran4}
\end{figure}

Beyond the average mutant frequency, the periodic migration also
induces a measurable phase lag between the mutant
frequency oscillations on the two islands. We quantify this lag by the phase
difference
\begin{equation}\label{equ:phase}
\delta_{\theta} = 2\pi \,\frac{\delta \tau}{T}, \qquad
\delta \tau = \bigl|t_{\max(n_1)} - t_{\max(n_2)}\bigr|,
\end{equation}
As shown in Fig.~\ref{fig:Moran2}, the vertical red and blue lines indicate the times $t_{\max(n_i)}$ at which the mutant frequency on island $i$ reaches its maximum. The horizontal separation between these lines quantifies the time delay $\delta \tau$ between the two populations.
Figure~\ref{fig:Moran4} shows~$\delta_\theta$ as a function of
migration amplitude~$\alpha_0$ and fitness
difference~$\delta r = r_{A1} - r_{A2}$ for three seasonal
periods, under condition~\eqref{equation1}.  The reason for employing condition \eqref{equation1} is that the phase quantity is only well-defined within this selection regime and these migration amplitudes . In the other regimes, extinction and fixation of the mutant type are observed , rendering the phase quantity undefined. As is evident in Figure~\ref{fig:Moran4}, analogous behavior is observed, across different migration scenarios over various parameter ranges. 
Several features are
apparent. First, the phase difference peaks at intermediate migration
amplitudes ($\alpha_0 \approx 0.4$--$0.5$), where the competing
tendencies of local selection and gene flow are most closely
balanced. At high migration amplitudes ($\alpha_0 \gtrsim 0.6$),
mutants rapidly colonize both islands, synchronizing the oscillations
and driving $\delta_\theta \to 0$. Second, $\delta_\theta$ increases
with the seasonal period~$T$: longer seasons allow the two island
populations to diverge more before the next migration pulse
resynchronizes them. Third, the phase difference is largely
insensitive to the fitness difference~$\delta r$, indicating that the
lag is set primarily by the migration parameters rather than by the
strength of local selection. Note that in the displayed symmetric
in-phase case, the lag is large at low migration amplitudes and decreases as the amplitude increases,
whereas the symmetric out-of-phase case generally exhibits smaller lags. In the asymmetric in-phase
case, the lag also depends on the fitness difference, particularly at larger periods.

We now turn to the regime in which the mutant is deleterious on
average across the two islands (Table \ref{tab:regimes}):
\begin{equation}\label{equation2}
\tag{II}
\frac{r_{A1} + r_{A2}}{2} < r_B , \quad r_{A1} > r_B , \quad
r_{A2} < r_B.
\end{equation}
Figure~\ref{fig:Moran3}(A,\,II) shows that constant migration
suppresses~$\bar\phi$ well below the isolation baseline: continuous
gene flow imports disfavored mutants into island~2 at a rate that
outweighs the benefit of exporting favored mutants from island~1.
Introducing periodicity partially offsets this
reduction. Even at moderate amplitudes, periodic migration raises~$\bar\phi$ substantially above the constant-migration value, partially offsetting the overall fitness disadvantage and enabling mutant recovery. The dependence on the seasonal period~$T$ is
non-monotonic and parameter-dependent, but longer periods generally
favor higher mutant frequencies. Under out-of-phase migration, the
rescue effect is even more pronounced, providing the strongest
recovery of mutant abundance among all symmetric scenarios,
particularly at long seasonal periods (\textcolor{blue}{Appendix C \ref{sec:S1.2} }).

In the final regime, the mutant is advantageous on average (Table \ref{tab:regimes}):
\begin{equation}\label{equation3}
\tag{III}
\frac{r_{A1} + r_{A2}}{2} > r_B , \quad r_{A1} > r_B , \quad
r_{A2} < r_B.
\end{equation}
Figure~\ref{fig:Moran3}(B,\,III) shows the out-of-phase case under
this regime. Constant migration elevates~$\bar\phi$ above the
isolation baseline, reflecting the net advantage of the mutant.
Periodic migration reduces~$\bar\phi$ relative to the
constant-migration limit, mirroring the pattern seen under
regime~\eqref{equation1} but with larger magnitude. The suppressive
effect of periodicity is strongest under out-of-phase migration and
intensifies markedly with increasing seasonal period~$T$. The
corresponding in-phase results (\textcolor{blue}{Appendix C \ref{sec:S1.3} }) show the same qualitative pattern but with a weaker reduction in~$\bar\phi$.

Taken together, the symmetric results reveal a consistent pattern:
periodic migration contracts~$\bar\phi$ toward the isolation
baseline. It reduces~$\bar\phi$ when the mutant is advantageous
(regimes~I and~III) and increases it when the mutant is deleterious
(regime~II), but never crosses the $\bar\phi = 0.5$ boundary. This
contraction toward neutrality is amplified by out-of-phase migration
and, in most parameter regimes, by longer seasonal periods.

\begin{figure}[htbp]
\hfill
\hspace*{-0.1\textwidth}
\includegraphics[width=1.1\textwidth]{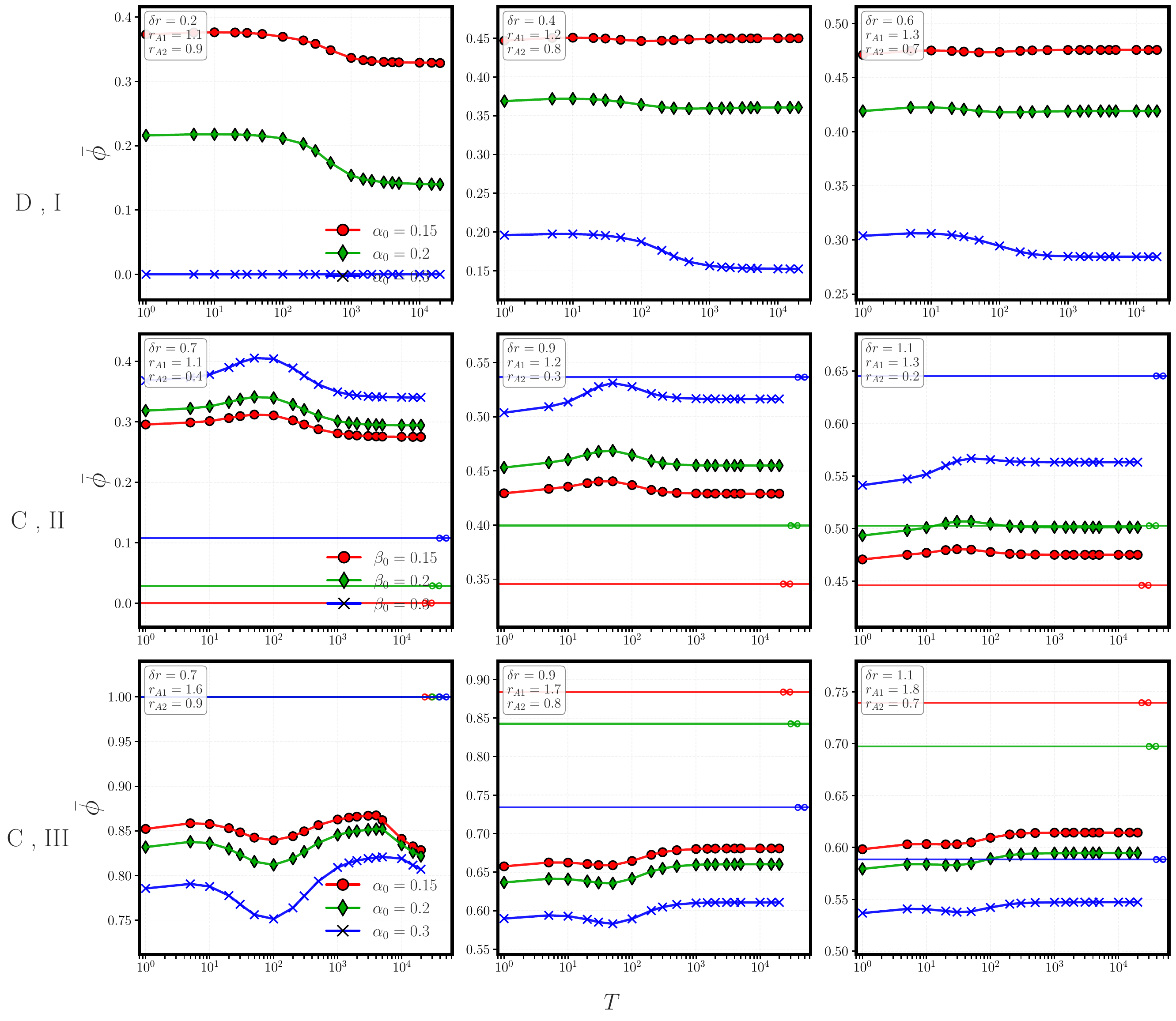}
\caption{Time-averaged mutant frequency~$\bar\phi$ as a function
of migration period~$T$ (logarithmic scale) under asymmetric
periodic migration. Rows correspond to three representative
scenario--regime combinations:
(D,\,I)~asymmetric out-of-phase with
condition~\eqref{equation1}, $\beta_0 = 0.1$,
$\delta r = 0.2$, $0.4$, $0.6$;
(C,\,II)~asymmetric in-phase with
condition~\eqref{equation2}, $\alpha_0 = 0.1$,
$\delta r = 0.7$, $0.9$, $1.1$;
(C,\,III)~asymmetric in-phase with
condition~\eqref{equation3}, $\beta_0 = 0.1$,
$\delta r = 0.7$, $0.9$, $1.1$.
Columns show increasing~$\delta r$ from left to right. In
panels~(D,\,I) and~(C,\,III): $\alpha_0 = 0.15$~($\circ$),
$\alpha_0 = 0.2$~($\diamond$), $\alpha_0 = 0.3$~($\times$); in
panel~(C,\,II): $\beta_0 = 0.15$~($\circ$),
$\beta_0 = 0.2$~($\diamond$), $\beta_0 = 0.3$~($\times$). Solid
lines with open symbols at $T \to \infty$ indicate the
large-period limit.}
\label{fig:Moran5}
\end{figure}

\subsection{Asymmetric Periodic Migration}
\label{sec:asymmetric}

We now allow the migration amplitudes in the two directions to
differ ($\alpha_0 \neq \beta_0$), so that gene flow has a
preferred direction. This introduces an additional degree of
freedom: whether the stronger flow is directed toward the island
where mutants are favored or toward the island where they are
disfavored. Combined with the in-phase versus out-of-phase
distinction, this yields a richer landscape of outcomes than the
symmetric case. A consistent feature of the asymmetric regime is
that the deviations between periodic and constant migration are
substantially larger than in the symmetric case: directional bias
amplifies the impact of temporal forcing.

Under neutral average selection~\eqref{equation1} with
preferential migration toward the island where mutants are
disfavored, Figure~\ref{fig:Moran5}(D,\,I) shows that mutant
abundance remains below the isolation baseline for all parameter
values examined. Increasing the seasonal period~$T$ consistently
decreases~$\bar\phi$, deepening the deficit. When the local fitness
difference is small and the migration amplitude is large,
$\bar\phi$~is nearly insensitive to~$T$: the system is close
enough to neutrality that temporal structure has little leverage.
The corresponding in-phase results (\textcolor{blue}{Appendix C \ref{sec:S2.1}}) show
the opposite trend --- longer periods increase~$\bar\phi$ ---
highlighting that the phase relationship between directional flows
can reverse the sign of the period dependence even when all other
parameters are held fixed. When migration is instead directed
toward the island of higher mutant fitness, the recovery
of~$\bar\phi$ is more pronounced, particularly at short seasonal
periods, and out-of-phase migration again produces the stronger
effect (\textcolor{blue}{Appendix C \ref{sec:S2.1}} ).

Under disadvantageous average selection~\eqref{equation2}, we
examine the in-phase regime with preferential migration toward the
island of higher mutant fitness.
Figure~\ref{fig:Moran5}(C,\,II) reveals one of the most striking
results of the study. At moderate to high migration amplitudes,
constant migration suppresses~$\bar\phi$ below the isolation
baseline; yet introducing periodicity drives a strong recovery,
raising~$\bar\phi$ substantially for periods up to
$T \approx 10^3$. At one particular combination of fitness
difference and amplitude (e.g.\ the $\delta r = 1.1$,
$\beta_0 = 0.2$ curve), the large-period limit value
of~$\bar\phi$ coincides exactly with the isolation baseline,
creating an effective neutral zone in which periodic and constant
migration yield identical steady states. Away from this zone, a
clear pattern emerges: when~$\bar\phi$ lies above the isolation
baseline, longer periods increase it further; when~$\bar\phi$ lies
below, longer periods decrease it. In the corresponding
out-of-phase regime, shorter periods almost universally
enhance~$\bar\phi$ across parameters. When migration instead
favors the island of lower mutant fitness, both in-phase and
out-of-phase behaviors are nearly indistinguishable and keep
$\bar\phi$ consistently below the isolation baseline
(\textcolor{blue}{Appendix C \ref{sec:S2.2} }).

Under advantageous average selection~\eqref{equation3}, we focus on
in-phase migration directed toward the island where mutants have
lower local fitness. Figure~\ref{fig:Moran5}(C,\,III) shows that,
because mutants are favored on average, constant migration
raises~$\bar\phi$ well above the isolation baseline, and in some
parameter regions mutants approach complete fixation. Periodic
migration reduces~$\bar\phi$ relative to the constant-migration
limit throughout, consistent with the contraction toward neutrality
observed in the symmetric case. A notable feature is that the curves have similar shapes as functions of T, although their levels depend on migration amplitude, 
%curves with the same local fitness difference~$\delta r$ behave nearly%
%identically regardless of migration amplitude%
indicating that local fitness asymmetry, rather than the details of gene flow,
is the dominant determinant of long-term dynamics in this regime.
The corresponding out-of-phase results are qualitatively similar
but exhibit a consistently stronger suppressive effect. When
migration is directed toward the island of higher mutant fitness,
the in-phase and out-of-phase behaviors converge: the large overall
fitness advantage of the mutant diminishes the distinction between
phase relationships (See \textcolor{blue}{Appendix C \ref{sec:S2.3} }).

\medskip

Several patterns emerge consistently across all symmetric and
asymmetric scenarios.

First, the dependence of~$\bar\phi$ on the seasonal period~$T$ is
generically non-monotonic. Rather than varying simply with season
length, mutant abundance exhibits turning points at intermediate
periods, reflecting an interplay between the timescale of migration
oscillations and the intrinsic relaxation timescale of local
selection. Under in-phase migration, this non-monotonicity gives
rise to a distinct optimal period~$T_{\mathrm{opt}}$ at
which~$\bar\phi$ is minimized in the advantageous regime and
maximized in the deleterious regime.

Second, out-of-phase migration consistently produces the most
extreme departures from the constant-migration baseline، whether
beneficial or detrimental to the mutant. The relative timing of
directional flows is thus a primary determinant of the magnitude of
the periodicity effect, often outweighing the influence of
amplitude or period length.

Third, the contraction toward neutrality is sign-preserving:
periodic migration reduces the effective strength of selection
but never inverts its direction. When mutants are advantageous on
average, periodicity attenuates their excess above $\bar\phi =
0.5$; when they are deleterious, it attenuates their deficit. The
modulation never crosses the neutrality line.

\subsection{Remarks on the Effects of Migration Periodicity}
\label{sec:remarks}

Several common patterns emerge across the migration scenarios considered
above. Periodic migration acts as a contraction toward neutrality: it
decreases $\bar\phi$ when the mutant is advantageous
($\bar\phi>0.5$) and increases $\bar\phi$ when the mutant is deleterious
($\bar\phi<0.5$), without inducing crossings of the neutrality threshold
$\bar\phi=0.5$ in the parameter regimes examined. This contraction is
strongest under out-of-phase migration, for which the attenuation toward
neutrality is most pronounced across both fitness regimes. Introducing
asymmetric migration further increases the disparity between periodic
and constant migration, indicating that directional asymmetry can amplify
the effect of temporal variation.

A second common feature is the non-monotonic dependence of $\bar\phi$ on
the migration period $T$. Under in-phase migration, an intermediate
optimal period $T_{\mathrm{opt}}$ emerges at which $\bar\phi$ is minimized
in the advantageous regime and maximized in the deleterious regime.
Under asymmetric migration, the migration amplitude also has a substantial
but fitness-regime-dependent influence on $\bar\phi$. Taken together,
these results show that populations with the same time-averaged migration
rate can exhibit different steady-state compositions depending on the
period, relative phase, and directional asymmetry of migration.

\section{Discussion}
\label{sec:discussion}

We have studied the effect of periodic, seasonally structured
migration on the steady-state mutant frequency in a two-deme Moran
death--birth process with spatially heterogeneous selection. Our
central result is that temporal structure in migration acts as a
\emph{contractor toward neutrality}: it attenuates the departure
of~$\bar\phi$ from the isolation baseline without inverting the
direction of selection. This contraction is sign-preserving:
advantageous mutants are suppressed and deleterious mutants are
rescued, but the modulation never crosses the $\bar\phi = 0.5$
neutrality line. The effect is amplified by out-of-phase migration,
by directional asymmetry, and is generically non-monotonic in the
seasonal period, with an intermediate optimal period~$T_{\mathrm{opt}}$
at which the modulation is extremal.

The contraction toward neutrality can be understood intuitively as
follows. Under constant migration, gene flow continuously couples
the two demes, allowing advantageous mutants on island~1 to
steadily colonize island~2 (or, in the deleterious case, allowing
the resident majority on island~2 to steadily erode the mutant
population on island~1). Periodic migration interrupts this
coupling: during the low-migration phase of each cycle, the two
islands evolve quasi-independently under local selection, which
drives each island toward its locally favored type. The net effect
of alternating between coupled and decoupled phases is to weaken
the time-averaged influence of gene flow relative to local
selection. Because gene flow is what pushes~$\bar\phi$ away from
the isolation baseline in both the advantageous and deleterious
regimes, weakening it contracts~$\bar\phi$ back toward that
baseline. The stronger contraction under out-of-phase migration is
consistent with this picture: when peak emigration from one island
coincides with minimal emigration from the other, the two
directional flows never reinforce each other simultaneously,
further reducing the effective coupling between demes.

The non-monotonic dependence on the seasonal period~$T$ suggests an
interplay between two timescales: the period of migration
oscillations and the intrinsic relaxation timescale of local
selection on each island. At very short periods ($T \to 0$), the
migration rate oscillates faster than the population can respond,
and the system effectively averages over the oscillation, behaving
as if migration were constant at its time-averaged rate. At very
long periods ($T \to \infty$), each half-cycle is long enough for
the population to relax fully to the instantaneous migration
conditions, and the system again approaches the constant-migration
limit. The strongest departure from the constant-migration
baseline occurs at intermediate periods where neither averaging nor
full relaxation applies, giving rise to the optimal
period~$T_{\mathrm{opt}}$. A quantitative characterization
of~$T_{\mathrm{opt}}$ in terms of model parameters, and whether
it can be predicted analytically from a separation-of-timescales
argument, remains an open question.

It is instructive to compare our results with the effects of
temporal fluctuations in \emph{selection} rather than migration.
Cvijovi\'c et al.~\cite{cvijovic2015fate} showed that temporal fluctuations in fitness
can substantially reduce the efficiency of selection and,
under sufficient variability, reverse the fate of a mutation
relative to its time-averaged fitness. Our findings share the
feature that temporal variation reduces the effective strength of
selection, but differ in a crucial respect: fluctuating migration
contracts toward neutrality without ever inverting the direction of
selection. This distinction likely reflects the different roles
that selection and migration play in the dynamics. Selection acts
directly on allele frequencies within each deme, so temporal
variation in selection can change the sign of the local fitness
difference and thereby reverse the direction of evolution.
Migration, by contrast, couples demes without directly altering
local fitness; temporal variation in migration modulates the
\emph{strength} of this coupling but cannot change the sign of
local selection, hence the sign-preserving character of the
contraction.

Our results also connect to earlier work on temporally variable
migration. Nagylaki~\cite{nagylaki1979island} investigated
allele-frequency dynamics under stochastic migration
(see also Ref.~\cite{rice2009evolution}), while Peniston
et al.~\cite{peniston2019pulsed} showed that pulsed immigration
can facilitate adaptation to harsh sink environments.
Our study complements this work by examining how the period,
relative phase, and directional asymmetry of periodic migration
affect long-term mutant abundance in two demes with fixed,
spatially heterogeneous selection.

An important mechanistic observation noted in Sec. 3.3 is that the 
direct effect of migration, the physical
transport of individuals between islands, is less consequential
for~$\bar\phi$ than the \emph{indirect} effect mediated through
local selection. Importing a mutant into its disfavored habitat
does not simply add one mutant to that island; it also displaces a
locally fitter resident, altering the competitive balance in
subsequent generations. Periodic migration modulates the strength
of this indirect coupling, and it is this modulation, rather than a
simple change in the number of migrants, that drives the
contraction toward neutrality. This observation has implications
for how migration is typically parametrized in population-genetic
models: the time-averaged migration rate alone may be an
insufficient summary statistic when migration is temporally
structured, because the indirect effects are nonlinear functions of
the instantaneous migration rate.

Several limitations of the present study suggest natural directions
for future work. First, our analysis is based on the deterministic
mean-field limit ($N \to \infty$); in finite populations, stochastic
fluctuations interact with periodic migration in ways that may
alter both the magnitude of the contraction toward neutrality and
the location of~$T_{\mathrm{opt}}$, and could additionally affect
fixation probabilities and times. Extending the present framework
to finite~$N$ using diffusion approximations or direct stochastic
simulation would clarify the robustness of our results to
demographic noise. Second, we have considered only two demes with a
single mutant locus. Natural metapopulations often comprise many
patches with complex connectivity networks, and multilocus dynamics
may introduce linkage effects that interact with periodic migration
in nontrivial ways. Third, the migration waveform in our model is
smooth and sinusoidal; natural migration events can be sharply
pulsed, asymmetric in waveform, or stochastic in timing. Whether
the contraction toward neutrality persists under more general
temporal migration profiles is an open question, though the
robustness of the effect across the four scenarios we examined
suggests it may be a generic consequence of temporal averaging
rather than an artifact of the specific waveform.

Finally, we have treated migration as a property of the environment, 
identical for mutants and residents, rather than as an
evolvable trait. In many natural systems, motility itself is under
selection, and the temporal structure of migration may coevolve with
local adaptation. Our results suggest that the timing of seasonal
migration, not only its magnitude, can be a target of natural
selection: organisms that time their dispersal to exploit the
contraction toward neutrality (for instance, by synchronizing
emigration with periods of locally unfavorable selection) could gain
a fitness advantage over those that migrate at constant rates. This
possibility connects our findings to the broader literature on the
evolution of dispersal timing and
phenology~\cite{krieger2017effects}.

The ecological relevance of our results extends to any system in
which populations occupying distinct selective environments are
connected by migration whose intensity varies periodically.
Migratory birds and mammals such as caribou that move annually between 
breeding and non-breeding ranges with different predation, disease, or
resource regimes represent a direct realization of the two-deme
framework with seasonally structured, potentially asymmetric gene
flow. Marine and island populations whose connectivity fluctuates
with tidal cycles, ocean currents, or sea-level
changes~\cite{princepe2024intermittent} experience periodic gene
flow on timescales ranging from days to millennia. In agricultural
and epidemiological settings, insect vectors and pathogen
populations disperse seasonally between refugia and crop or host
habitats that impose different selection on resistance alleles, and
our results suggest that the seasonal timing of this dispersal may
be as consequential for the persistence of locally adapted variants
as its average rate. More broadly, our findings underscore that the
temporal pattern of migration is not a second-order correction to
its time-average but can qualitatively reshape the balance between
local adaptation and gene flow, a factor that deserves explicit
attention in both theoretical models and empirical studies of
spatially structured evolution.

\section*{Data Availability}
All Python codes used in this study are available in the GitHub repository at: \href{https://github.com/Amin-physics1380/Seasonal_Migration}{https://github.com/Amin-physics1380/Seasonal Migration}.

\section*{Acknowledgement}

The authors thank Melika Gorgi for insightful comments and constructive discussions. KK acknowledges the hospitality of the Abdus Salam International Centre for Theoretical Physics (ICTP) during a visit. MRE acknowledges the hospitality of Sultan Qaboos University during a visit. MRE further acknowledges support from ICTP through the Sabbatical Programme and from the Simons Foundation (Record ID: SFI-MPS-T-Institutes-00012057; AD), which also supported this work.

\section*{Appendices}
\appendix
\setcounter{equation}{0}
\renewcommand{\theequation}{A\arabic{equation}}
\section{Mean-field limit of the Moran process}
\label{app:meanfield}
The stochastic dynamics of the mutant count~$n_i$ on island~$i$
are governed by a master equation. Let $p(n_i,t)$ denote the
probability that island~$i$ contains $n_i$ mutants at time~$t$.
Probability conservation under the death--birth transitions gives
\begin{align}
\label{eq:master}
\frac{dp(n_i,t)}{dt}
  &= W^+(n_i-1)\,p(n_i-1,t)
  + W^-(n_i+1)\,p(n_i+1,t)\nonumber\\
  &- \bigl[W^+(n_i) + W^-(n_i)\bigr]\,p(n_i,t),
\end{align}
where $W^+(n_i)$ and $W^-(n_i)$ are the transition probabilities
for increasing and decreasing the mutant count by one,
respectively~\cite{moran, nowak2006evolutionary}. The mean mutant
count $\langle n_i \rangle = \sum_{n_i=0}^{N} n_i\, p(n_i,t)$
evolves according to
\begin{equation}
\label{eq:exact-mean}
\frac{d\langle n_i\rangle}{dt}
  = \langle W^+(n_i)\rangle - \langle W^-(n_i)\rangle,
\end{equation}
obtained by multiplying~\eqref{eq:master} by~$n_i$ and summing
over all states. This equation is exact but not closed, since the
right-hand side involves averages of nonlinear functions of~$n_i$.
In the limit of large deme size ($N \gg 1$), fluctuations about
the mean become negligible and one may approximate
$\langle W^\pm(n_i)\rangle \approx W^\pm(\langle n_i\rangle)$.
Defining the relative mutant frequency
$\phi_i = \langle n_i \rangle / N$, this mean-field closure yields
the deterministic equation
\begin{equation}
\label{eq:meanfield}
\frac{d\phi_i}{dt}
  = W^+(\phi_i) - W^-(\phi_i),
\end{equation}
which governs the mutant frequency dynamics throughout the main
text.
For a two-island model , it can be written : 
\begin{equation}
\frac{d\phi_i}{dt}
  = W^+_i(\phi_1 , \phi_2 ) - W^-_i(\phi_1 , \phi_2),
\end{equation}
The explicit forms of~$W^\pm_i$ for the two-island
death--birth model with migration are derived in
Appendix~\ref{app:transitions}.

\setcounter{equation}{0}
\renewcommand{\theequation}{B\arabic{equation}}
\section{Transition probabilities for the two-island death-birth model}
\label{app:transitions}
We derive the transition probabilities for the two-island system
under the death-birth update rule with migration. At each step, a
focal island is chosen and one of its residents is selected
uniformly at random for removal. A replacement is then produced by
a parent drawn, with probability proportional to fitness, from a
pool that includes individuals on the focal island (with weight
$\mu_{ii}$) and individuals on the other island (with weight
$\mu_{ji}$). Whether the new offspring is a mutant or a resident
therefore depends on the mutant frequencies on both islands and on
the instantaneous migration probabilities.
For island~1, the probability that the mutant count increases by
one is
\begin{align}
\label{eq:Wplus1}
W^+_{DB,1}(\phi_1, \phi_2)
  &= \underbrace{%
       \frac{(1 - \phi_1)\, d_B}
            {\phi_1 d_A + (1-\phi_1) d_B
             + \phi_2 d_A + (1-\phi_2) d_B}
     }_{\text{a resident on island 1 is removed}}
     \notag \\[6pt]
  &\quad \times
     \underbrace{%
       \frac{\mu_{11}\,\phi_1\, r_{A1}
             + \mu_{12}\,\phi_2\, r_{A2}}
            {\mu_{11}\,\phi_1\, r_{A1}
             + \mu_{12}\,\phi_2\, r_{A2}
             + (1 - \mu_{11}\,\phi_1
                  - \mu_{12}\,\phi_2)\, r_B}
     }_{\text{a mutant is chosen as parent}}.
\end{align}
The first factor is the probability that the individual selected
for death on island~1 is a resident (type~$B$), weighted by its
death rate~$d_B$. The second factor is the probability that the
replacement offspring is a mutant, given that the parent is drawn
from island~1 (with probability~$\mu_{11}$, contributing mutants
at fitness~$r_{A1}$) or from island~2 (with
probability~$\mu_{12}$, contributing mutants at
fitness~$r_{A2}$).
The probability that the mutant count on island~1 decreases by one
is
\begin{align}
\label{eq:Wminus1}
W^-_{DB,1}(\phi_1, \phi_2)
  &= \underbrace{%
       \frac{\phi_1\, d_A}
            {\phi_1 d_A + (1-\phi_1) d_B
             + \phi_2 d_A + (1-\phi_2) d_B}
     }_{\text{a mutant on island 1 is removed}}
     \notag \\[6pt]
  &\quad \times
     \underbrace{%
       \frac{(1 - \mu_{11}\,\phi_1
                - \mu_{12}\,\phi_2)\, r_B}
            {\mu_{11}\,\phi_1\, r_{A1}
             + \mu_{12}\,\phi_2\, r_{A2}
             + (1 - \mu_{11}\,\phi_1
                  - \mu_{12}\,\phi_2)\, r_B}
     }_{\text{a resident is chosen as parent}}.
\end{align}
The mean-field equation for island~1 is then
\begin{equation}
\label{eq:dphi1}
\frac{d\phi_1}{dt}
  = W^+_{DB,1}(\phi_1,\phi_2)
  - W^-_{DB,1}(\phi_1,\phi_2).
\end{equation}
The transition probabilities for island~2 are obtained by
exchanging the island indices ($1 \leftrightarrow 2$) throughout:
\begin{align}
\label{eq:Wplus2}
W^+_{DB,2}(\phi_2, \phi_1)
  &= \frac{(1 - \phi_2)\, d_B}
          {\phi_1 d_A + (1-\phi_1) d_B
           + \phi_2 d_A + (1-\phi_2) d_B}
     \notag \\[6pt]
  &\quad \times
     \frac{\mu_{22}\,\phi_2\, r_{A2}
           + \mu_{21}\,\phi_1\, r_{A1}}
          {\mu_{22}\,\phi_2\, r_{A2}
           + \mu_{21}\,\phi_1\, r_{A1}
           + (1 - \mu_{22}\,\phi_2
                - \mu_{21}\,\phi_1)\, r_B},
     \\[8pt]
\label{eq:Wminus2}
W^-_{DB,2}(\phi_2, \phi_1)
  &= \frac{\phi_2\, d_A}
          {\phi_1 d_A + (1-\phi_1) d_B
           + \phi_2 d_A + (1-\phi_2) d_B}
     \notag \\[6pt]
  &\quad \times
     \frac{(1 - \mu_{22}\,\phi_2
              - \mu_{21}\,\phi_1)\, r_B}
          {\mu_{22}\,\phi_2\, r_{A2}
           + \mu_{21}\,\phi_1\, r_{A1}
           + (1 - \mu_{22}\,\phi_2
                - \mu_{21}\,\phi_1)\, r_B},
\end{align}
with
\begin{equation}
\label{eq:dphi2}
\frac{d\phi_2}{dt}
  = W^+_{DB,2}(\phi_2,\phi_1)
  - W^-_{DB,2}(\phi_2,\phi_1).
\end{equation}
The migration probabilities satisfy the normalization
\begin{equation}
\label{eq:mu-norm}
\mu_{11} + \mu_{12} = 1, \qquad
\mu_{22} + \mu_{21} = 1,
\end{equation}
and the specific periodic forms of $\mu_{12}(t)$ and $\mu_{21}(t)$
for each scenario are given in Table~\ref{tab:scenarios}.
Equations~\eqref{eq:dphi1} and~\eqref{eq:dphi2}, together with
the time-dependent migration rates, form a non-autonomous coupled
ODE system that is integrated numerically to obtain the results in
the main text. 
%In our research we set $d_A = d_B = 1$ and $r_B = 1$ which this choice reflects the assumption that the death rates of resident and mutant individuals are identical (there is no difference in mortality between the two types), and that the resident type has the same fitness in both habitats%.
In this study , we set $d_A = d_B =1$ and $r_B =1$, assuming equal mortality for both types and equal resident birth rates in the two habitats. Only the mutant type exhibits different fitnesses on the two islands. 

\setcounter{equation}{0}
\renewcommand{\theequation}{C\arabic{equation}}
\section{Supplementary results and figures}
%\label{sec:SM}% 
\noindent
This section provides the additional results that were omitted from the main text. We systematically examine the remaining combinations of migration scenarios (symmetric and asymmetric, in-phase and out-of-phase) and selective regimes (I--III). Throughout, the time-averaged migration rates are kept constant, allowing us to isolate the effect of the temporal organization of gene flow on the steady-state mutant frequency $\bar{\phi}$. The results confirm and extend the main conclusion of the paper: periodic migration acts as a sign-preserving contraction toward neutrality.
% ============================================================
\subsection{Symmetric periodic migration --- remaining combinations}
%\label{sec:S1}%
% ============================================================
In the symmetric cases the two directional amplitudes are equal ($\alpha_0 = \beta_0$). Consequently, the time-averaged migration rate in each direction is fixed at $\overline{\mu}_{12} = \overline{\mu}_{21} = \alpha_0/2$ for small periods. Below we report the three scenario--regime combinations omitted from Figure~\ref{fig:Moran3} of the main text.
\subsubsection{Symmetric out-of-phase migration under regime I}
\label{sec:S1.1}
\begin{figure}[htbp]
\centering
\includegraphics[width=\textwidth,keepaspectratio]{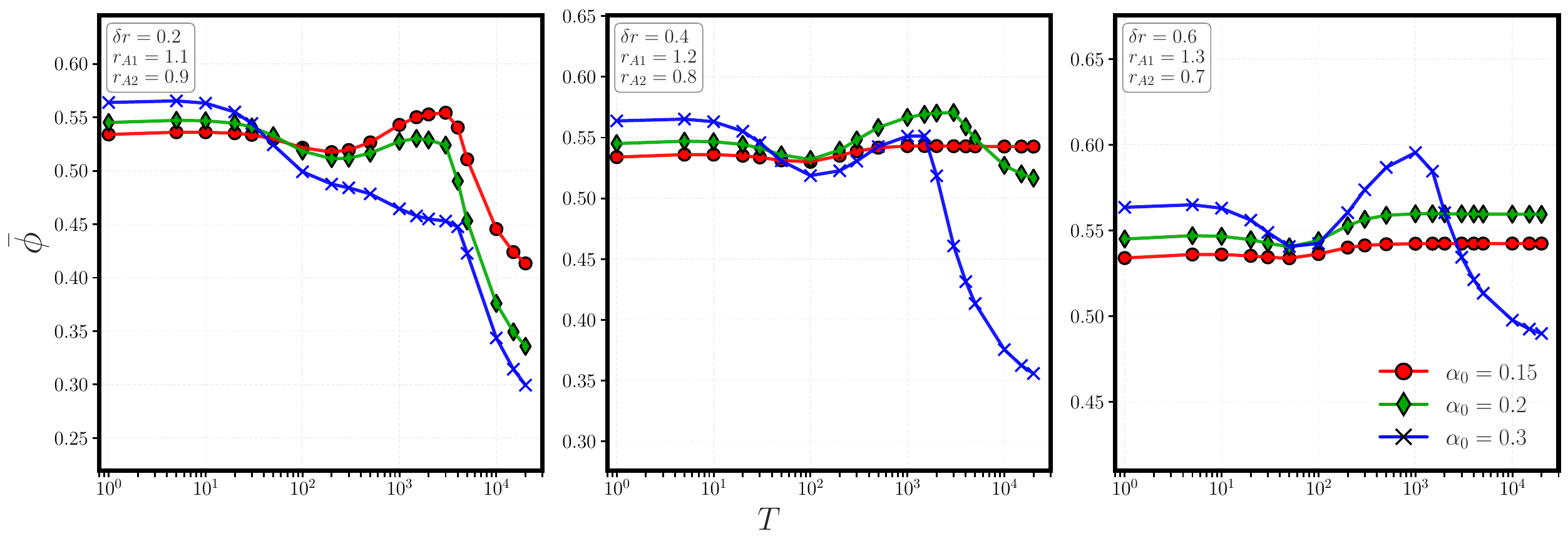}
\caption{Time-averaged mutant frequency $\bar{\phi}$ as a function of migration period $T$ (logarithmic scale) under symmetric out-of-phase migration (Scenario~B) and neutral average selection (regime~I). Different symbols correspond to different migration amplitudes $\alpha_0$. Solid lines with open symbols at $T\to\infty$ indicate the large-period limit.}
\label{fig:S1}
\end{figure}

Figure~\ref{fig:S1} shows the time-averaged mutant frequency $\bar{\phi}$ as a function of the migration period $T$ under symmetric out-of-phase migration (Scenario~B) and neutral average selection (regime~I). Constant migration elevates $\bar{\phi}$ above the isolation baseline $\bar{\phi} = 0.5$. When the two directional flows are offset by half a period, the reduction of $\bar{\phi}$ relative to this baseline is substantially stronger than under the corresponding in-phase case (Figure~\ref{fig:Moran3}(A,I) of the main text). The magnitude of the suppression increases with $T$. Across the full range of amplitudes examined, $\bar{\phi}$ remains above the isolation baseline yet approaches it less closely than under in-phase migration. The dependence on $T$ is only non-monotonic.

\subsubsection{Symmetric out-of-phase migration under regime II}
\label{sec:S1.2}
\begin{figure}[htbp]
\centering
\includegraphics[width=\textwidth,keepaspectratio]{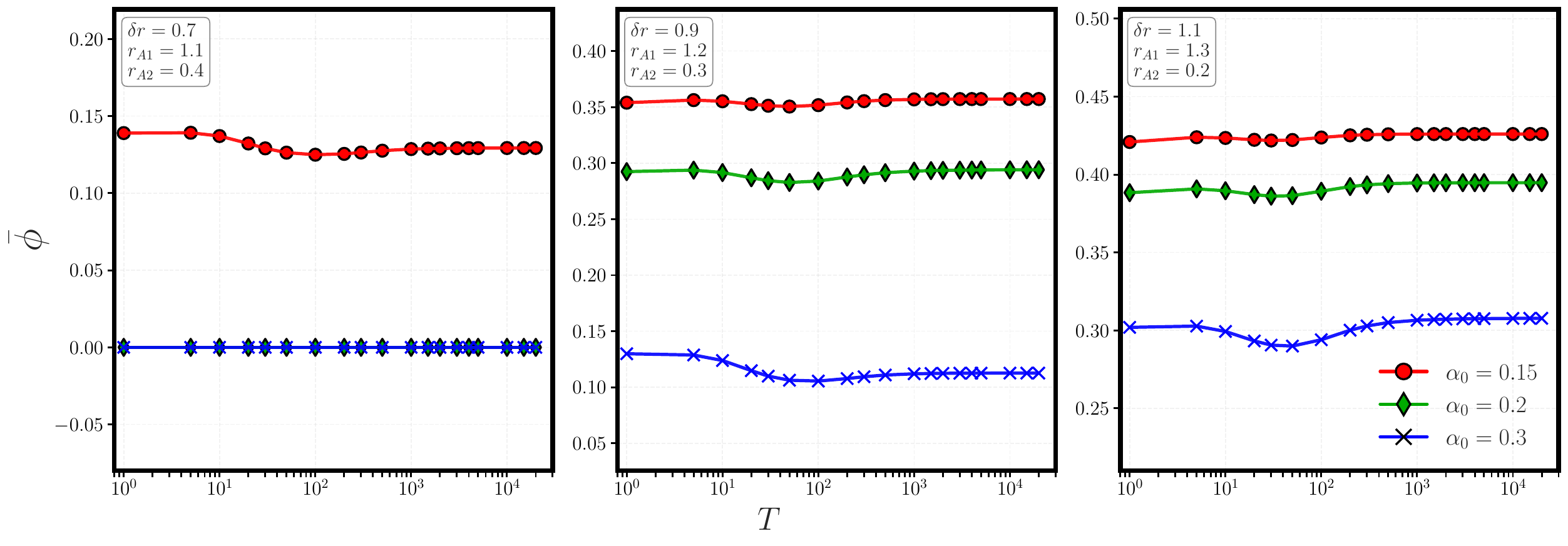}
\caption{Time-averaged mutant frequency $\bar{\phi}$ as a function of migration period $T$ under symmetric out-of-phase migration (Scenario~B) and deleterious average selection (regime~II). Different symbols correspond to different amplitudes $\alpha_0$. Solid lines mark the large-period limit.}
\label{fig:S2}
\end{figure}

Figure~\ref{fig:S2} shows results for symmetric out-of-phase migration under deleterious average selection (regime~II). Constant migration suppresses $\bar{\phi}$ well below the isolation baseline. When the directional flows are out of phase, the elevation of $\bar{\phi}$ above this baseline is the strongest among all symmetric cases examined. The rescue effect is particularly pronounced at long seasonal periods and, for small amplitudes, can bring $\bar{\phi}$ close to (but still below) the isolation baseline.

\subsubsection{Symmetric in-phase migration under regime III}
\label{sec:S1.3}
\begin{figure}[htbp]
\centering
\includegraphics[width=\textwidth,keepaspectratio]{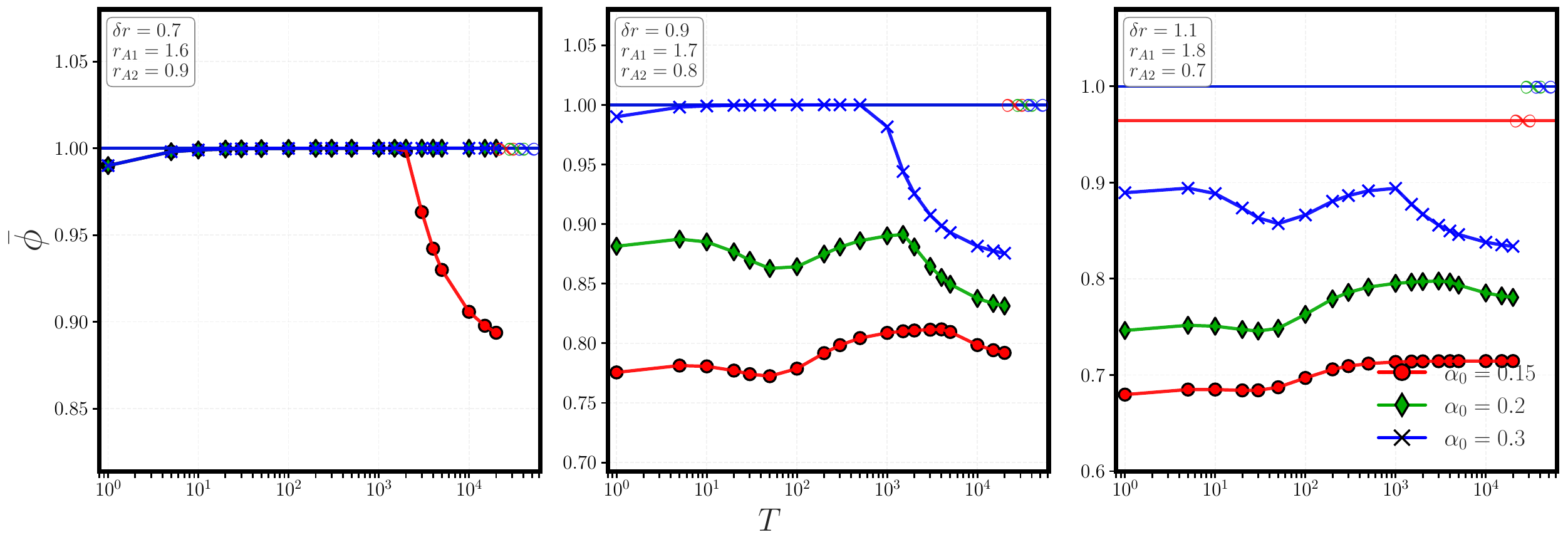}
\caption{Time-averaged mutant frequency $\bar{\phi}$ as a function of migration period $T$ under symmetric in-phase migration (Scenario~A) and advantageous average selection (regime~III). Different symbols correspond to different amplitudes $\alpha_0$. Solid lines mark the large-period limit.}
\label{fig:S3}
\end{figure}

Figure~\ref{fig:S3} shows the corresponding results for symmetric in-phase migration (Scenario~A) under advantageous average selection (regime~III). Constant migration raises $\bar{\phi}$ well above the isolation baseline. Periodic modulation reduces $\bar{\phi}$ relative to this limit, although the reduction is weaker than under the corresponding out-of-phase case (Figure~\ref{fig:Moran3}(B,III) of the main text), especially for longer periods. The dependence on $T$ is non-monotonic and exhibits a shallow minimum at intermediate periods. As is evident, $\bar{\phi}$ changes across different periods.

\subsection{Asymmetric periodic migration --- remaining combinations}
%\label{sec:S2}%

In the asymmetric cases the two directional amplitudes are independent ($\alpha_0 \neq \beta_0$), introducing a preferred direction of gene flow. The combinations omitted from Figure~5 of the main text are reported below.
\subsubsection{Asymmetric migration under regime I}
\label{sec:S2.1}
\begin{figure}[htbp]
\centering
\includegraphics[width=\textwidth,keepaspectratio]{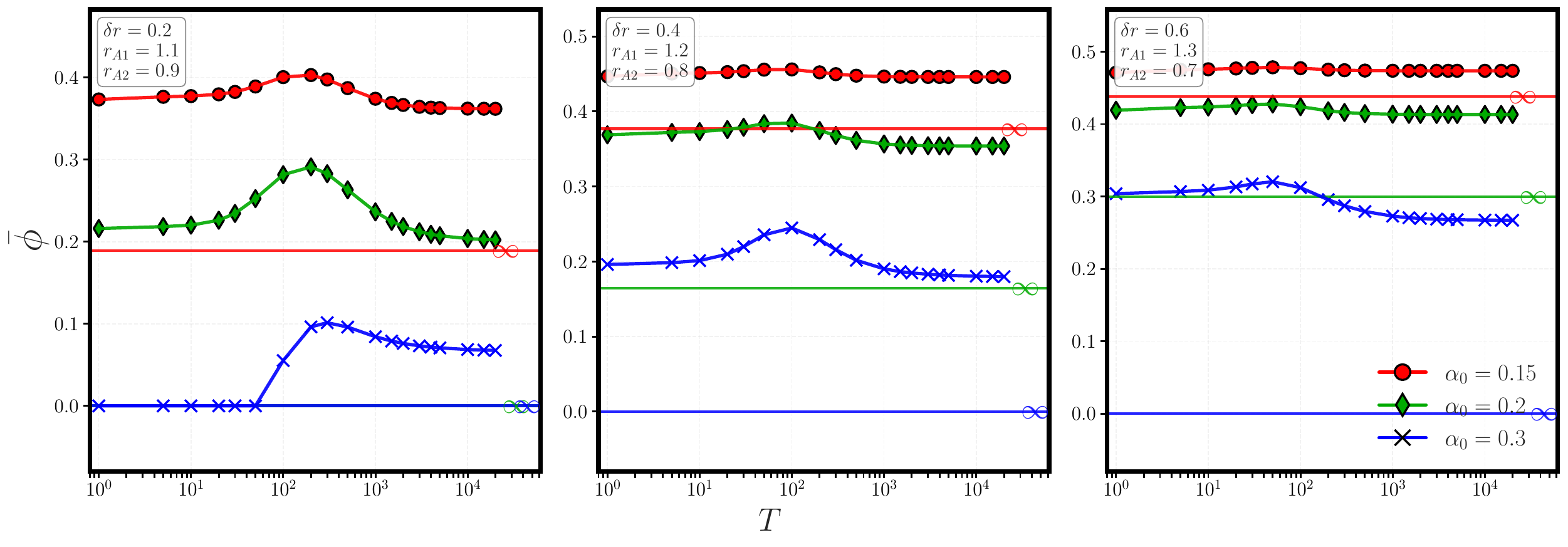}\\[1.2ex]
\includegraphics[width=\textwidth,keepaspectratio]{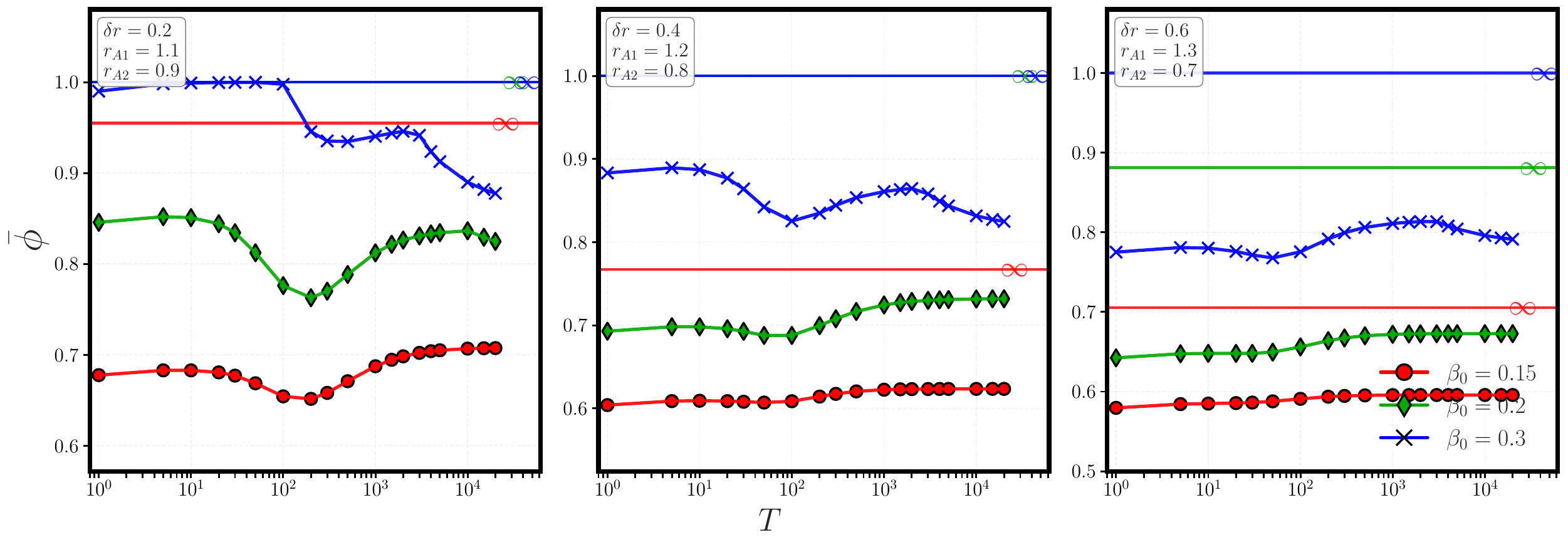}\\[1.2ex]
\includegraphics[width=\textwidth,keepaspectratio]{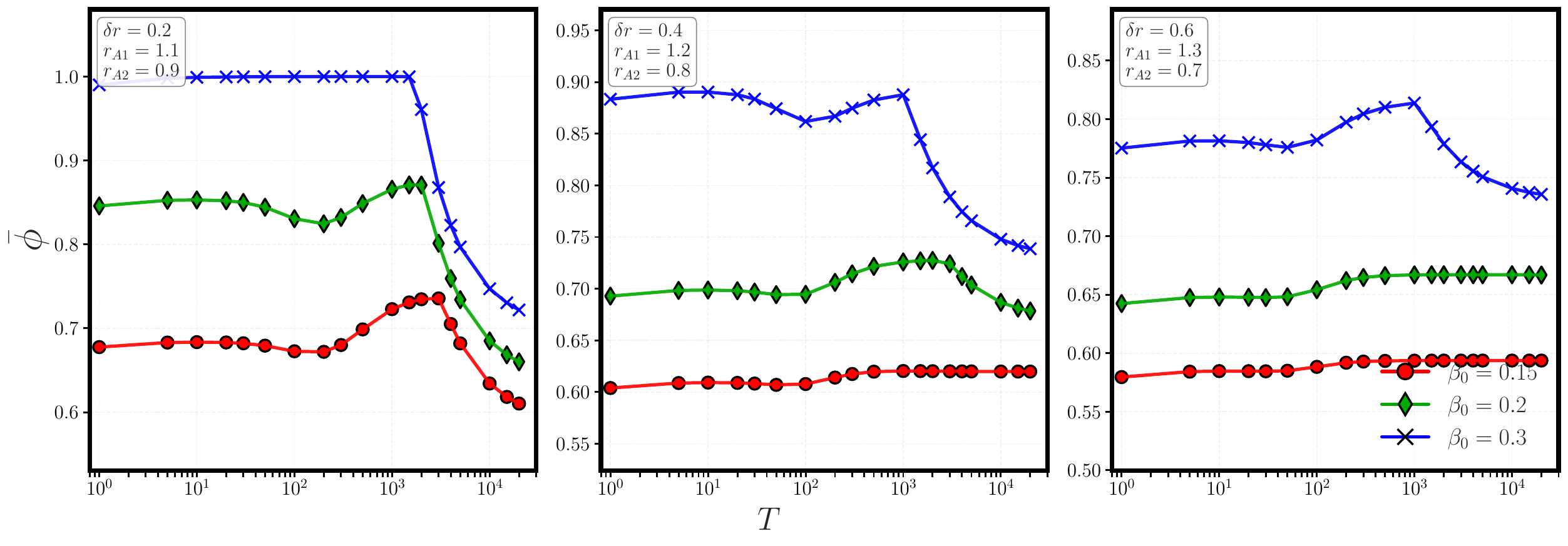}
\caption{Time-averaged mutant frequency $\bar{\phi}$ versus migration period $T$ under asymmetric migration and neutral average selection (regime~I). 
\textbf{Top:} asymmetric in-phase with fixed $\beta_0$ = 0.1.
\textbf{Middle:} asymmetric in-phase with fixed $\alpha_0$ = 0.1. 
\textbf{Bottom:} asymmetric out-of-phase with fixed $\alpha_0$ = 0.1. 
Solid lines indicate the large-period limits.}
\label{fig:S4}
\end{figure}

Figure~\ref{fig:S4} shows $\bar{\phi}$ versus $T$ under asymmetric migration and neutral average selection (regime~I). When preferential migration is directed toward the island where mutants are locally disfavoured and the flows are in phase (Scenario~C), longer periods increase $\bar{\phi}$. This trend is opposite to that observed for the corresponding out-of-phase case (Figure~\ref{fig:Moran5}(D,I) of the main text). The elevation remains modest and $\bar{\phi}$ stays below the isolation baseline. When the stronger flow is instead directed toward the higher-fitness island, the recovery of $\bar{\phi}$ is more pronounced, particularly at short periods, and is again stronger under out-of-phase timing for longer periods.
\begin{figure}[htbp]
\centering
\includegraphics[width=\textwidth,height=0.27\textheight,keepaspectratio]{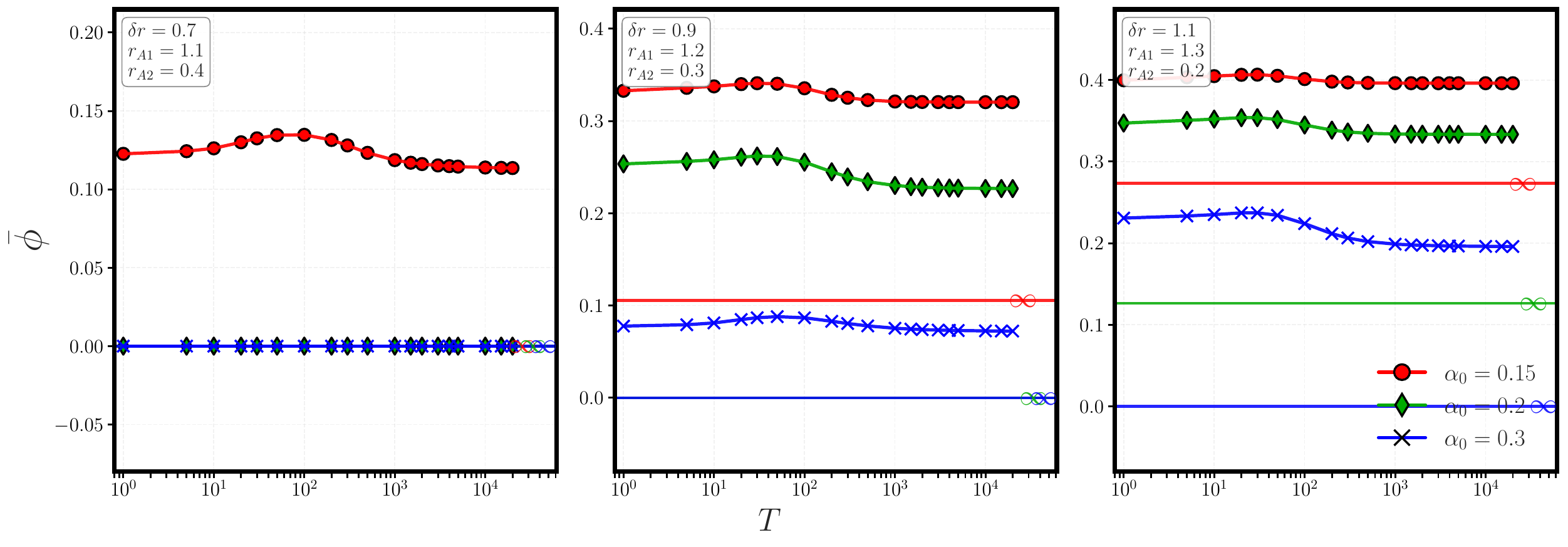}\\[0.8ex]
\includegraphics[width=\textwidth,height=0.27\textheight,keepaspectratio]{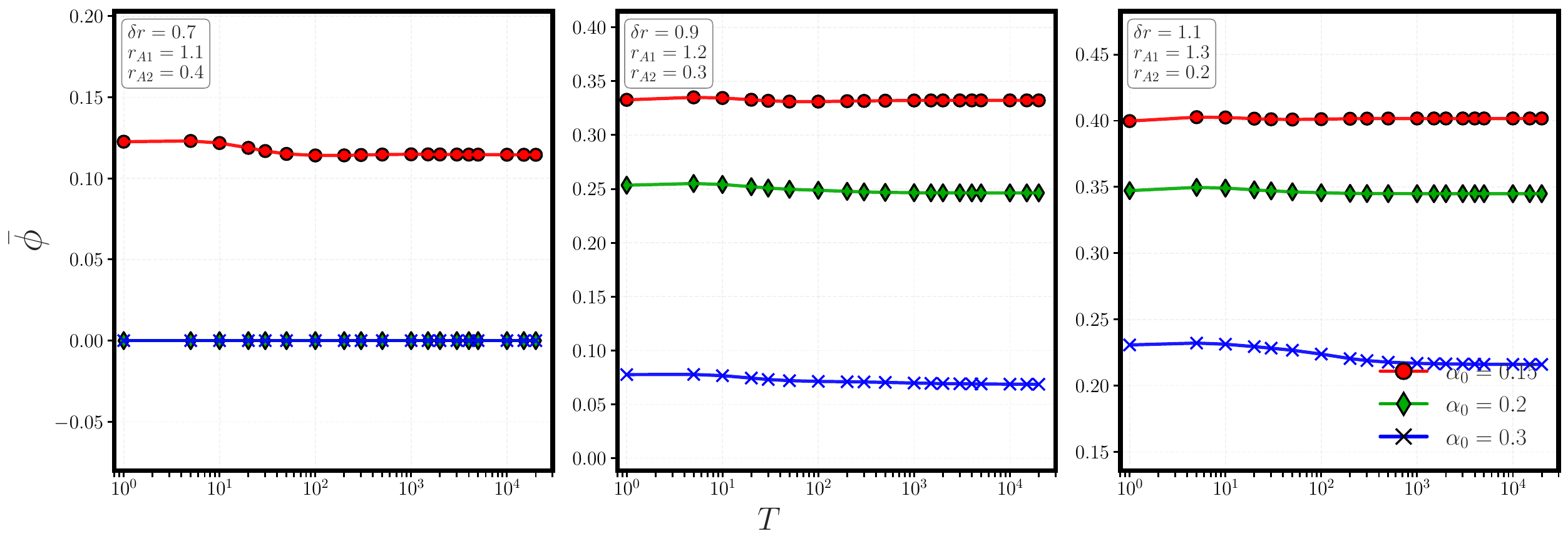}\\[0.8ex]
\includegraphics[width=\textwidth,height=0.27\textheight,keepaspectratio]{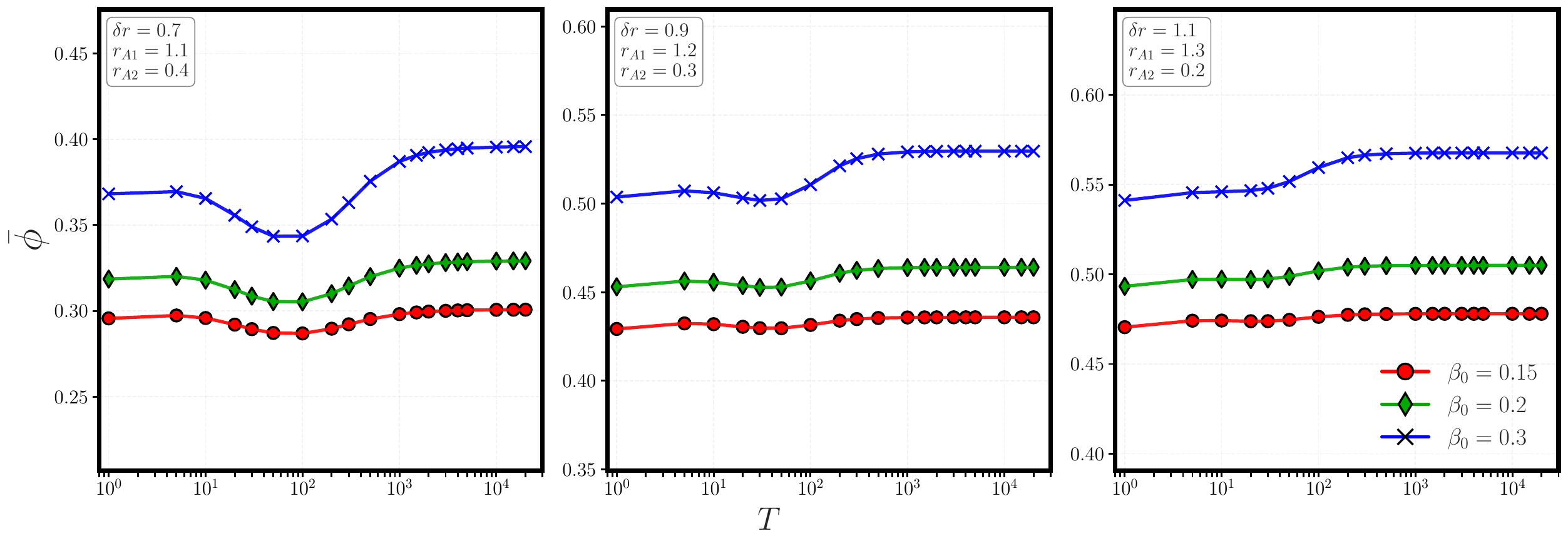}
\caption{Time-averaged mutant frequency $\bar{\phi}$ versus migration period $T$ under asymmetric migration and deleterious average selection (regime~II). 
\textbf{Top:} asymmetric in-phase with fixed $\beta_0 = 0.1$. 
\textbf{Middle:} asymmetric out-of-phase with fixed $\beta_0 = 0.1$. 
\textbf{Bottom:} asymmetric out-of-phase with fixed $\alpha_0  =0.1$. 
Solid lines mark the large-period limits.}
\label{fig:S5}
\end{figure}
\begin{figure}[htbp]
\centering
\includegraphics[width=\textwidth,height=0.27\textheight,keepaspectratio]{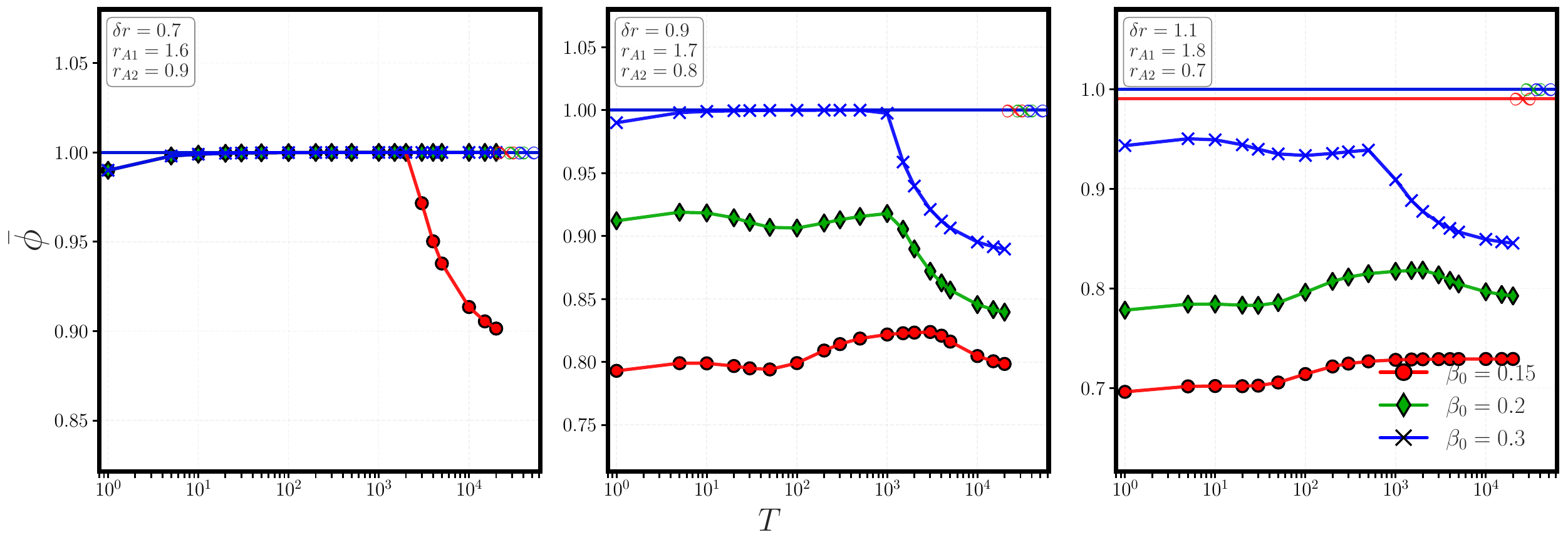}\\[0.8ex]
\includegraphics[width=\textwidth,height=0.27\textheight,keepaspectratio]{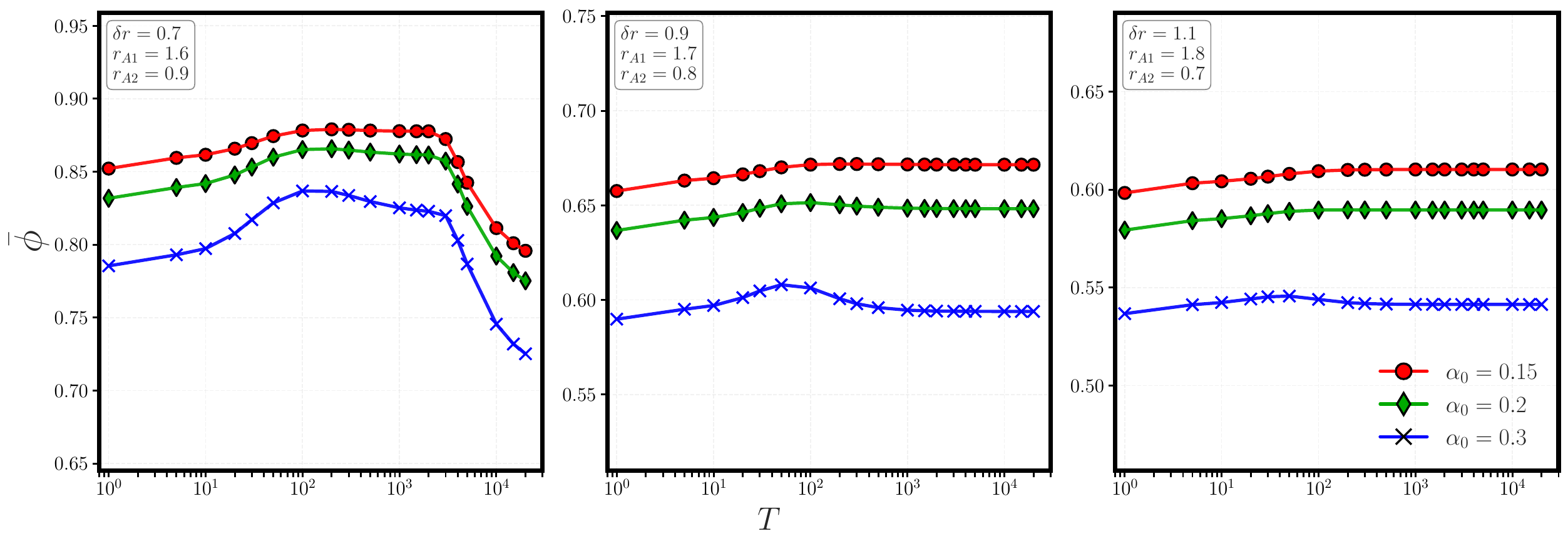}\\[0.8ex]
\includegraphics[width=\textwidth,height=0.27\textheight,keepaspectratio]{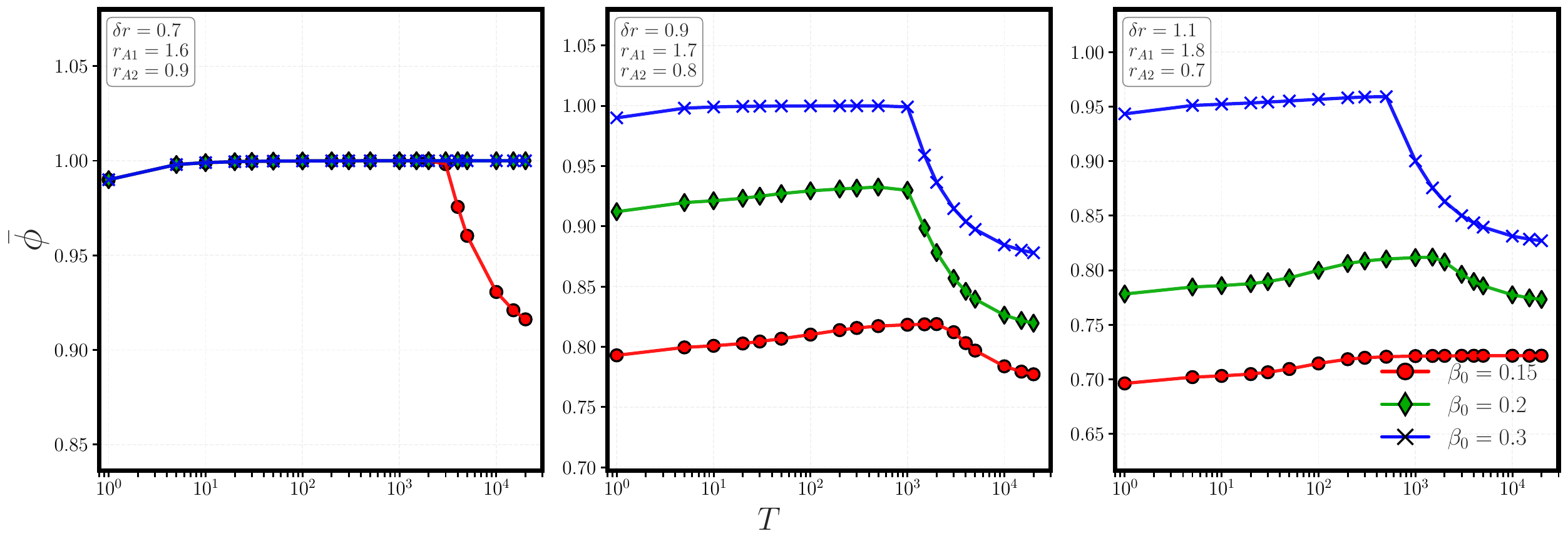}
\caption{Time-averaged mutant frequency $\bar{\phi}$ versus migration period $T$ under asymmetric migration and advantageous average selection (regime~III). 
\textbf{Top:} asymmetric in-phase with fixed $\alpha_0$ = 0.1. 
\textbf{Middle:} asymmetric out-of-phase with fixed $\beta_0$ = 0.1. 
\textbf{Bottom:} asymmetric out-of-phase with fixed $\alpha_0$ = 0.1. 
Solid lines mark the large-period limits.}
\label{fig:S6}
\end{figure}
\subsubsection{Asymmetric migration under regime II}
\label{sec:S2.2}
Figure~\ref{fig:S5} shows results under deleterious average selection (regime~II). When preferential migration is directed toward the higher-fitness island and the flows are out of phase (Scenario~D), shorter periods almost universally enhance $\bar{\phi}$. The recovery relative to the constant-migration baseline is strong and can exceed that obtained in the corresponding in-phase case (Figure~\ref{fig:Moran5}(C,II) of the main text). When migration is preferentially directed toward the lower-fitness island, both in-phase and out-of-phase cases keep $\bar{\phi}$ consistently below the isolation baseline and are nearly indistinguishable from each other.
\subsubsection{Asymmetric migration under regime III}
\label{sec:S2.3}
Figure~\ref{fig:S6} shows results under advantageous average selection (regime~III). Out-of-phase migration produces a consistently stronger suppression of $\bar{\phi}$ than the corresponding in-phase case (Figure~\ref{fig:Moran5}(C,III) of the main text). The shapes of the $\bar{\phi}$ curves remain similar across amplitudes, indicating that local fitness asymmetry is the dominant determinant of the long-term dynamics once the mutant enjoys a net selective advantage. When the stronger flow is directed toward the higher-fitness island, the distinction between in-phase and out-of-phase timing largely disappears.
\vspace{1.2em}

\noindent
Taken together, the results presented in this Supplementary Material confirm that periodic migration acts as a robust, sign-preserving contraction toward neutrality across all examined scenario--regime combinations. The effect is consistently strongest under out-of-phase timing and is further amplified by directional asymmetry.
\printbibliography

\end{document}